\documentclass[11pt,a4paper]{article}
\pdfoutput=1

\usepackage{jheppub}
\usepackage[rgb]{xcolor}
\usepackage{color}
\usepackage{amsmath,amsfonts,mathrsfs,graphicx,bbm,dsfont,booktabs,mathtools,braket,lmodern,tikz}

\DeclareMathAlphabet{\mathpzc}{OT1}{pzc}{m}{it}

\def\L{{\mathbb L}}

\newcommand{\bea}{\begin{eqnarray}}
\newcommand{\eea}{\end{eqnarray}}
\def\be{\begin{equation}}
\def\ee{\end{equation}}
\newcommand{\bei}{\begin{itemize}}
\newcommand{\eei}{\end{itemize}}
\newcommand{\bee}{\begin{enumerate}}
\newcommand{\eee}{\end{enumerate}}

\def\a {\alpha}
\def\b {\beta}

\def\p{\phi}

\def\ov{\over}

\def\pa {\partial}

\def\L{\mathscr L}

\def\ads{{\rm AdS}_5\times {\rm S}^5}

\def\ads{{\rm AdS}_5\times {\rm S}^5}

\def\am{{\rm am}}
\def\am0{{\rm am}_0}

\expandafter\def\expandafter\bfseries\expandafter{\bfseries\ifmmode\else\boldmath\fi}
\expandafter\def\expandafter\mdseries\expandafter{\mdseries\ifmmode\else\unboldmath\fi}
\expandafter\def\expandafter\normalfont\expandafter{\normalfont\ifmmode\else\unboldmath\fi}

\definecolor{grey}{rgb}{0.4,0.4,0.5}
\definecolor{darkgreen}{rgb}{0,0.5,0}
\definecolor{darkred}{rgb}{0.6,0.0,0}
\definecolor{lightbrown}{rgb}{1,0.9,0.8}
\definecolor{brown}{rgb}{0.6,0.3,0.3}
\definecolor{darkblue}{rgb}{0,0,0.8}
\definecolor{darkmagenta}{rgb}{0.5,0,0.5}

\title{Integrability of the $\eta$-deformed\\
 Neumann-Rosochatius model}

\author[a,1]{Gleb Arutyunov}
\note{Correspondent fellow at Steklov Mathematical Institute, Moscow.}
\author[a]{Martin Heinze}
\author[b]{and Daniel Medina-Rincon}

\affiliation[a]{Institut f\"ur Theoretische Physik, Universit\"at Hamburg, Luruper Chaussee 149, 22761 Hamburg, Germany}
\affiliation[a]{Zentrum f\"ur Mathematische Physik, Universit\"at Hamburg, Bundesstrasse 55, 20146 Hamburg, Germany}

\affiliation[b]{
Nordita, KTH Royal Institute of Technology and Stockholm University, Roslagstullsbacken 23, SE-106 91 Stockholm, Sweden}
\affiliation[b]{Department of Physics and Astronomy, Uppsala University SE-751 08 Uppsala, Sweden}

\emailAdd{gleb.arutyunov@desy.de}
\emailAdd{martin.heinze@desy.de}
\emailAdd{d.r.medinarincon@nordita.org}

\abstract{An integrable deformation of the well-known Neumann-Rosochatius system is studied by considering generalised bosonic spinning solutions on the $\eta$-deformed $\rm{AdS}_{5}\times\rm{S}^{5}$ background. For this integrable model we construct a $4\times 4$ Lax representation and a set of integrals of motion that ensures its Liouville integrability.  These integrals of motion correspond to the deformed analogues of the Neumann-Rosochatius integrals and generalise the previously found integrals for the $\eta$-deformed Neumann and $(\rm{AdS}_{5}\times\rm{S}^{5})_{\eta}$ geodesic systems. Finally, we briefly comment on consistent truncations of this model.}

\begin{document}

\begin{flushright}\small{NORDITA-2016-84\\ZMP-HH/16-14\\UUTP-13/16}\end{flushright}

\maketitle


\section{Introduction}
The usage of integrable techniques in the context of the gauge-string correspondence \cite{M}  provided us with an unpreceded analytic insight 
into the problem of  higher orders of perturbation theory in the planar maximally supersymmetric ${\cal N}=4$ gauge theory, as well as with important results for finite values of the 
coupling and even for non-perturbative ones, see {\it e.g.} \cite{Arutyunov:2009ga, Beisert:2010jr} for the reviews.  Most of our progress and understanding came through investigation of the light-cone string sigma model 
on $\ads$ by means of the Factorised Scattering Theory, Thermodynamic Bethe Ansatz  (TBA)  \cite{Arutyunov:2009ur}-\cite{Gromov:2009bc}  and the quintessence of the latter realised in the form of the quantum spectral curve 
construction \cite{Gromov:2013pga}.  In many cases this progress was possible due to ingenious guesswork, an intuition developed in studying a number of simpler examples, through comparisons 
with different limiting cases where a solution was possible by other means and also by trial and error.  To better understand the nature of the proposed constructions, their analytic properties and the role of symmetries, it is essential to study other solvable examples of stringy type sigma models and their gauge theory duals.  One such interesting example  is offered by integrable deformations of the $\ads$ string sigma model \cite{Delduc:2013qra,Delduc:2014kha} based on the earlier constructions by Klimcik \cite{Klimcik:2002zj, Klimcik:2008eq}.
In modern language these deformations can be classified as $\eta$-deformations \cite{Delduc:2013qra}, $\lambda$-deformations \cite{Sfetsos:2013wia,Hollowood:2014qma} and deformations related to solutions of the classical Yang-Baxter equation   
\cite{Kawaguchi:2014qwa}-\cite{vanTongeren:2015soa}.  In some cases these deformations are not totally unrelated but can be connected through the contraction limits \cite{Hoare:2016hwh,Hoare:2016ibq}.

\smallskip

In the present paper we restrict our attention to certain aspects of the $\eta$-deformed sigma model on $\ads$. We recall that our current knowledge about this model includes the perturbative S-matrix \cite{Arutyunov:2013ega,Arutyunov:2015qva}, which agrees with the semi-classical 
limit of the exact S-matrix based on the quantum group symmetries \cite{BK}. Assuming that the exact S-matrix  \cite{BK} drives the scattering in the quantum version of the $\eta$-deformed model, in a series of works   \cite{Arutynov:2014ota}-\cite{vanTongeren:2013gva} the mirror TBA construction\footnote{For the construction of the $q$-deformed dressing phase see \cite{Hoare:2011wr}.} for the corresponding sigma model spectrum has been developed. Importantly, as was recently shown  \cite{Arutyunov:2015mqj}, the target-space bosonic fields (NSNS and RR in the string theory language)  of the two-dimensional $\eta$-deformed model 
do not satisfy the standard type IIB supergravity equations but rather obey their specific generalisation, an observation also confirmed by considerations of classical $\kappa$-symmetry \cite{Wulff:2016tju}. A similar phenomenon has been also observed for other deformations 
\cite{Hoare:2016hwh,Orlando:2016qqu,Kyono:2016jqy}.

\smallskip

As is well known, under various consistent reductions an integrable two-dimensional sigma model can produce one-dimensional integrable models which might have an important physical meaning on their own.
For instance, for the $\ads$ sigma model the spinning string ansatz produces rigid string solutions \cite{Frolov:2002av,Frolov:2003qc},  which are nicely described in terms of the Neumann or Neumann-Rosochatius models \cite{Arutyunov:2003uj,Arutyunov:2003za}.
The integrable models of Neumann or Neumann-Rosochatius type are historically among the first examples of integrable systems and they show up in various problems of mathematical physics including the problem of geodesics on ellipsoid 
or equivariant harmonic maps into spheres \cite{Neumann}-\cite{JMoser}. In the context of strings on $\ads$ they were useful to explicitly construct the corresponding spinning string solutions and compute their energy as a function of spins.
The associated conserved charges governing the corresponding string profiles were compared to that of the spin chain solutions describing certain operators in ${\cal N}=4$ theory \cite{Arutyunov:2003rg}.

\smallskip

In the previous work by two of us \cite{Arutyunov:2014cda}, the deformed Neumann model which follows from the reduction of the $\eta$-deformed sigma model on $\ads$ was studied. 
There, without loss of generality  we restricted ourselves to the case of the deformed sphere and found the corresponding Lax connection and the deformed analogues of Uhlenbeck integrals. The aim of the present paper 
is to extend our previous analysis to the deformation of the Neumann-Rosochatius model, which also naturally comes from the $\eta$-deformed sigma model. The Neumann-Rosochatius model
is richer in the sense that its reduction to the ``Rosochatius part" describes the geodesic problem on a (deformed) sphere.  Again, under a certain limiting procedure we will extract the integrals of motion 
of the deformed Neumann-Rosochatius system from the Lax matrix of the $\eta$-deformed sigma model. We then explicitly demonstrate that these integrals are in involution with respect 
to the Dirac bracket. The three conserved angular momenta corresponding to the isometry directions together with two Neumann-Rosochatius integrals are enough to 
declare that the deformed model is integrable in the Liouville sense. We hope that our present findings on the integrability of the  deformed Neumann-Rosochatius system will be further used in explicit constructions
of corresponding solutions, see {\it e.g.}   \cite{Kameyama:2014vma}-\cite{Banerjee:2016xbb}, which are necessary to be able to compare with the results based on the exact TBA approach.

\smallskip

The paper is organised as follows. In the next section we recall the basic facts about the usual Neumann-Rosochatius system. In Section 3 we briefly describe the $\eta$-deformed sigma model and the spinning ansatz for the corresponding solutions.
Section 4 is devoted to the Lax representation for the $\eta$-deformed Neumann-Rosochatius model and integrals of motion. In Section 5 we continue the study of the integrals of motion, establish connection to 
the previously found integrability of the $\eta$-deformed Neumann and Rosochatius model, and discuss consistent truncations to lower-dimensional models. In the Conclusions we discuss the results obtained and formulate some open problems. Some technical details are collected in 3 appendices.


\section{The Neumann-Rosochatius model}\label{UndefSystems}

We will briefly review the undeformed integrable system, making special emphasis on its integrals of motion and the properties thereof. We start with presenting the main features of the Neumann-Rosochatius model, which describes, in particular, generalised spinning strings in $\ads$. Later, we will describe the relation between the Neumann-Rosochatius system and the Neumann and Rosochatius integrable models.

In the present section, all expressions will be given in the $x_{i}$ coordinates commonly used in the literature. The equivalent expressions in terms of the unconstrained coordinates $(r,\xi)$ are given in appendix \ref{undefinrxi}.

\subsection{Brief overview}\label{briefoverviewundef}

The Lagrangian for this system is given by
\begin{equation}\label{undefLagNr}
L_{NR}  = \frac{1}{2}\sum\limits_{i = 1}^3 {\left( {{x'_i}^2 + x_i^2{\alpha'_{i}}^2 - \omega _i^2x_i^2} \right) + \frac{\Lambda }{2}\left( {\sum\limits_{i = 1}^3 {x_i^2}  - 1} \right)} ,
\end{equation}
where $\Lambda$ is a Lagrangian multiplier and $'$ denotes a derivative with respect to time. The corresponding Hamiltonian is given by
\begin{equation}\label{HNR}
{H_{NR}} = \frac{1}{2}\sum\limits_{i = 1}^3 {\left( {\pi _i^2 + \omega _i^2x_i^2 + \frac{{\pi _{{\alpha _i}}^2}}{{x_i^2}}} \right)}=\frac{1}{4}\sum\limits_{i \ne j}^3 {J_{ij}^2}  + \frac{1}{2}\sum\limits_{i = 1}^3 {\left( {\omega _i^2x_i^2 + \frac{{\pi _{{\alpha _i}}^2}}{{x_i^2}}} \right)} \ ,
\end{equation}
where ${J_{ij}} = {x_i}{\pi _j} - {x_j}{\pi _i}$, with $\pi_{i}$ and $\pi _{{\alpha _i}}$ denoting the momenta canonically conjugate to $x_{i}$ and $\alpha_{i}$, respectively, and the Hamiltonian is subjected to the constraints
\begin{align}\label{undefconst}
 \sum\limits_{i = 1}^3 {x_i^2 = 1}\ ,&& \sum\limits_{i = 1}^3 {{x_i}{\pi _i} = 0}\ .
 \end{align}
Due to the second constraint, in these coordinates it is necessary to use the Dirac bracket formalism, which yields
\begin{align*}
{\left\{ {{\pi _i},{\pi _j}} \right\}_{D.B.}} = {x_i}{\pi _j} - {x_j}{\pi _i}\ ,&&{\left\{ {{\pi _i},{x_j}} \right\}_{D.B.}} = {\delta _{ij}} - {x_i}{x_j}\ ,&& {\left\{ {{x_i},{x_j}} \right\}_{D.B.}} = 0\ .
\end{align*}
From the Hamiltonian of this system, we see that it describes a particle moving on a sphere under both a harmonic oscillator potential with frequency $\omega_{i}$ (which in this context is also referred as the Neumann potential) and a Coulomb potential (also referred as the Rosochatius potential).
 
We have  three coordinates $x_{i}$, three phases $\alpha_{i}$, their corresponding conjugated momenta and two constraints. Thus, Liouville integrability requires 5 integrals of motion in involution in order to solve this system. Since the phases $\alpha_{i}$ are cyclic coordinates, their canonically conjugate momenta $\pi_{\alpha_i}$ are integrals of motion. As the three $\pi_{\alpha_i}$ are pairwise in involution, we can fix them to be constants, while $x_i$ and their conjugate momenta will describe the dynamics of the system.

The other two integrals of motion needed for complete integrability of the system can be chosen from the generalisation of the Uhlenbeck integrals
 \begin{equation}\label{IntNR}
 {I_i} = x_i^2 + \sum\limits_{j \ne i}^3 {\frac{1}{{\omega _i^2 - \omega _j^2}}} \left[ {J_{ij}^2 + \frac{{\pi _{{\alpha _i}}^2x_j^2}}{{x_i^2}} + \frac{{\pi _{{\alpha _j}}^2x_i^2}}{{x_j^2}}} \right]\quad i\in\left\{ {1,2,3} \right\}\, .
 \end{equation}
The integrals of motion satisfy the following properties
 \begin{equation}\label{propertiesundefNR1}
\left\{ {{H_{NR}},{I_i}} \right\}_{D.B.} = 0\ ,\quad \quad\quad\left\{ {{I_i},{I_j}} \right\}_{D.B.} = 0\ ,\quad \quad \quad \left\{ {{I_i},{\pi _{{\alpha _j}}}} \right\}_{D.B.} = 0\ ,
\end{equation}
\begin{equation}\label{propertiesundefNR2}
\sum\limits_{i = 1}^3 {{I_i} = 1} .
\end{equation}
From the last expression it is clear that only two of the three integrals $I_{i}$ are independent.
Furthermore, also the Hamiltonian can be expressed as a linear combination of the integrals of motion introduced above
\begin{equation}\label{propertiesundefNR3}
{H_{NR}} = \frac{1}{2}\sum\limits_{i = 1}^3 {\left( {\omega _i^2{I_i} + \pi _{{\alpha _i}}^2} \right)} .
\end{equation}

\subsection{Connection with the Neumann and Rosochatius integrable models}\label{connectionundef}
As previously mentioned, the Neumann-Rosochatius system is of particular interest since it has both the potential terms of the Neumann and Rosochatius integrable models. We will now briefly explain how the Neumann and Rosochatius integrable models are recovered as limits of the more general Neumann-Rosochatius system.

First, we will show how the Neumann-Rosochatius system reduces to the Neumann model in the limit of $\pi_{\alpha_i}\rightarrow0$. From equation \eqref{HNR} we see that $H_{N}$, the Hamiltonian of the Neumann model, is given by
\begin{align}\label{counterpart1a}
{H_N} = \mathop {\lim }\limits_{{\pi _{{\alpha _i}}} \to 0} {H_{NR}} = \frac{1}{2}\sum\limits_{i = 1}^3 {\left( {\pi _i^2 + \omega _i^2x_i^2} \right)} \ ,
\end{align}
still subjected to the constraints of equation \eqref{undefconst}. Naturally, in this limit, the integrals of motion of the Neumann-Rosochatius system reduce to the Uhlenbeck integrals of the Neumann model
\begin{align}\label{counterpart1b}
{F_i} = \mathop {\lim }\limits_{{\pi _{{\alpha _j}}} \to 0} {I_i} = x_i^2 + \sum\limits_{j \ne i}^3 {\frac{{J_{ij}^2}}{{\omega _i^2 - \omega _j^2}}} \quad\quad i \in \left\{ {1,2,3} \right\}\ .
\end{align}

Reduction of the Neumann-Rosochatius model to the Rosochatius system happens in the limit $\omega_{i}\rightarrow0$, where the Hamiltonian of the Rosochatius system is given by
\begin{equation}\label{HNRtoHR}
{H_R} = \mathop {\lim }\limits_{{\omega _j} \to 0} {H_{NR}} = \frac{1}{2}\sum\limits_{i = 1}^3 {\left( {\pi _i^2 + \frac{{\pi _{{\alpha _i}}^2}}{{x_i^2}}} \right)} \ ,
\end{equation}
subjected to the standard constraints \eqref{undefconst}. The integrals of motion for this system are given by
\begin{align}\label{IntR}
{F_{ij}} = J_{ij}^2 + \frac{{\pi _{{\alpha _i}}^2x_j^2}}{{x_i^2}} + \frac{{\pi _{{\alpha _j}}^2x_i^2}}{{x_j^2}}\quad i\neq j\ .
\end{align}
The integrals of motion of the Neumann-Rosochatius and Rosochatius systems are intrinsically related, which can be seen from expressions \eqref{IntNR} and \eqref{IntR}
$${I_i} = x_i^2 + \sum\limits_{j \ne i}^3 {\frac{{{F_{ij}}}}{{\omega _i^2 - \omega _j^2}}}\ .$$
Namely, taking into account the fact that $0\le x_{i}^2\le1$, we see that as the frequencies $\omega_{i}$, and by this $\omega_{i}^{2}-\omega_{j}^{2}$, approach zero, the leading contribution to the integrals $I_{i}$ will come from the Rosochatius integrals of motion $F_{ij}$,
\begin{equation}\label{NRtoNUndefLimit}
\mathop {\lim }\limits_{{\omega _{j}\to\omega_{i}}} \left( {\omega _i^2 - \omega _j^2} \right){I_i} = {F_{ij}}\ ,
\end{equation}
where $i\neq j$ and there is no summation over the $i$ index in the expression above.


\section{Bosonic $\eta$-deformed sigma model and generalised spinning solutions}
\label{section3}
Our starting point to analise the $\eta$-deformed Neumann-Rosochatius model is the Lagrangian of the bosonic $(\ads)_{\eta}$ sigma model \cite{Arutyunov:2013ega} restricted to the sphere
\bea
\L &=&{1\ov2}\, \eta^{\a\b}\Bigg(\frac{\pa_\a r\pa_\b r
  }{ \left(1-r^2\right) \left(1+\varkappa ^2 r^2\right)}
  +\frac{r^2 \pa_\a \xi\pa_\b \xi}{1+ \varkappa ^2 r^4 \sin ^2\xi}+\frac{r^2 \cos ^2\xi\ \pa_\a \p_1\pa_\b \p_1}{1+ \varkappa ^2
   r^4 \sin ^2\xi }\label{L} \nonumber \\
   &&\qquad +r^2 \sin^2\xi\ \pa_\a \p_2\pa_\b \p_2 + \frac{\left(1-r^2\right)\pa_\a \p_{3}\pa_\b \p_{3}}{1+\varkappa ^2 r^2}\, \Bigg)\label{FulletaLag} \\
   &&+{\varkappa\ov2}\epsilon^{\alpha\beta}\left(\frac{ r^4 \sin 2 \xi }{1+ \varkappa ^2 r^4 \sin^2\xi}\pa_\a\p_1\pa_\b\xi +\frac{2r \partial_{\alpha}r\partial_{\beta}\phi_{3}}{1+\varkappa^{2}r^{2}} \right)\ ,\nonumber
\eea
where $r\in\left[0,1\right]$, $\xi\in\left[0,\frac{\pi}{2}\right]$, $\phi_{i}\in\left[0,2\pi\right]$ and $\varkappa=\frac{2\eta}{1-\eta^2}\ge0$, where $\eta$ is the original deformation parameter of \cite{Delduc:2013qra}. For convenience, we choose the world-sheet metric $\eta^{\alpha\beta}$ to be Minkowski, while $\epsilon^{\alpha\beta}$ denotes the Levi-Civita symbol. Just as done in \cite{Arutyunov:2014cda}, we rescaled the Lagrangian by an overall constant and included a total derivative which appeared naturally in the calculation of the B-field contribution \cite{Arutyunov:2013ega}.

\begin{figure}[h]
\centering
\includegraphics [scale=0.60]{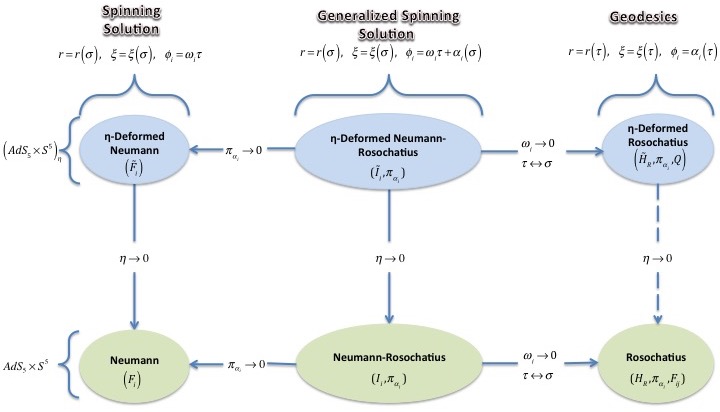}
\caption{Classical solutions in $\rm{AdS}_{5}\times \rm{S}^{5}$ and $(\rm{AdS}_{5}\times \rm{S}^{5})_{\eta}$ with their corresponding integrable models and integrals of motion.}
\label{noche12}
\end{figure}

In general, we will consider closed solutions along the world-sheet spatial direction $\sigma$ and consequently, we assume $r$ and $\xi$ to be periodic in $\sigma$ with period $2\pi$. From equation \eqref{FulletaLag} we see that $\L$ has three isometries corresponding to translations along the angles $\phi_{i}$. Thus, we can consider classical solutions located at a point in the center of $\rm{AdS}$ and having the following form on the $\eta$-deformed $\rm{S}^5$:
\begin{equation}
\begin{aligned}
\label{solucion2}
r= r(\sigma)\ ,&& \xi =\xi(\sigma)\ ,  \\
\phi_{1}=\omega_{1}\tau+\alpha_{1}(\sigma)\ ,\quad\quad\quad&& \phi_{2}=\omega_{2}\tau+\alpha_{2}(\sigma)\ ,\quad\quad\quad && \phi_{3}=\omega_{3}\tau+\alpha_{3}(\sigma)\ ,
\end{aligned}
\end{equation}
where $\tau$ and $\sigma$ denote the world-sheet coordinates, $\omega_{i}$ are constant angular velocities and $\alpha_{i}(\sigma)$ are interpreted as real phases satisfying the periodicity condition $\alpha_{i}(\sigma+2\pi)=\alpha_{i}(\sigma)+2\pi m_{i}$ with $m_{i}\in\mathbb{Z}$.

This ansatz represents a generalisation of the spinning solutions studied in \cite{Arutyunov:2014cda} (for which $\alpha_{i}\rightarrow0$) and therefore, we will refer to solutions of the form \eqref{solucion2} as ``generalised spinning solutions''. For the case of undeformed $\ads$, in \cite{Arutyunov:2003za} solutions of this type were shown  to reduce to a 1-d integrable model: The well-known Neumann-Rosochatius system, with $\sigma$ playing the role of time parameter, while $\tau$ decouples from the equations of motion. In the case of $\eta$-deformed $\ads$, this type of solution was first considered in \cite{Kameyama:2014vma}, where the Lagrangian was presented in a new set of coordinates. 

For generalised spinning solutions, the reduction of the Lagrangian \eqref{FulletaLag} and its corresponding Hamiltonian are given by
\begin{align}\label{LagNR}
{\widetilde L_{NR}} =& \frac{1}{2}\left[ \frac{{{{r'}^2}}}{{\left( {1 - {r^2}} \right)\left( {1 + {\varkappa ^2}{r^2}} \right)}} + \frac{{{r^2}{{\xi '}^2} + \varkappa {\omega _1}{r^4}\xi '\sin 2\xi }}{{1 + {\varkappa ^2}{r^4}{{\sin }^2}\xi }} + \frac{{\left( {{\alpha'}_1^2 - \omega _1^2} \right){r^2}{{\cos }^2}\xi }}{{1 + {\varkappa ^2}{r^4}{{\sin }^2}\xi }}\right.\\
&\left. + \left( {{\alpha '}_2^2 - \omega _2^2} \right){r^2}{{\sin }^2}\xi  + \frac{{\left( {{\alpha '}_3^2 - \omega _3^2} \right)\left( {1 - {r^2}} \right)}}{{1 + {\varkappa ^2}{r^2}}} - \frac{{2\varkappa {\omega _3}rr'}}{{1 + {\varkappa ^2}{r^2}}} \right]\nonumber\ ,\\
{\widetilde H_{NR}} = &\frac{1}{2}\left[ \left( {1 - {r^2}} \right)\left( {1 + {\varkappa ^2}{r^2}} \right)\pi _r^2 + \frac{{\pi _\xi ^2\left( {1 + {\varkappa ^2}{r^4}{{\sin }^2}\xi } \right)}}{{{r^2}}} - 2\varkappa {\omega _1}{\pi _\xi }{r^2}\sin \xi \cos \xi\right.\label{HamNrrr}\\
&\left. + 2\varkappa {\omega _3}{\pi _r}r\left( {1 - {r^2}} \right) + \omega _1^2{r^2}{{\cos }^2}\xi  + \omega _2^2{r^2}{{\sin }^2}\xi  + \omega _3^2\left( {1 - {r^2}} \right)\right.\nonumber\\
&\left. + \frac{{\pi _{{\alpha _1}}^2\left( {1 + {\varkappa ^2}{r^4}{{\sin }^2}\xi } \right)}}{{{r^2}{{\cos }^2}\xi }} + \frac{{\pi _{{\alpha _2}}^2}}{{{r^2}{{\sin }^2}\xi }} + \frac{{\pi _{{\alpha _3}}^2\left( {1 + {\varkappa ^2}{r^2}} \right)}}{{1 - {r^2}}} \right]\nonumber\ ,
\end{align}
where $\sigma$ plays the role of the time parameter and $'$ denotes $\partial_{\sigma}$. Again, by cyclicity of the $\alpha_i$,  the angular momenta $\pi_{\alpha_{i}}$ conjugate to $\alpha_{i}$ are integrals of motion. Since these three momenta are pairwise in involution, to have Liouville integrability one needs two extra conserved quantities in involution. These will be constructed in Section \ref{construyendo} (in principle one of them can be chosen to be $\widetilde{H}_{NR}$).

To see that this system indeed corresponds to a one-parameter deformation of the Neumann-Rosochatius model, one can easily check that moving from unconstrained coordinates $(r,\xi)$ to constrained coordinates $x_{i}$ given by:
\begin{align}\label{cocord}
x_{1}=r \cos\xi\ ,&& x_{2}=r\sin\xi\ ,&& x_{3}=\sqrt{1-r^2}\ ,
\end{align}
the Lagrangian \eqref{LagNR} constitutes a deformation of \eqref{undefLagNr} with the constraint $\sum_{i}x_{i}^{2}=1$.

As previously mentioned, the generalised spinning solution \eqref{solucion2} reduces to the spinning solution considered in \cite{Arutyunov:2014cda} by sending $\alpha_{i}\rightarrow0$ and $\pi_{\alpha_{i}}\rightarrow0$. Taking this limit in equations \eqref{LagNR} and \eqref{HamNrrr}, one obtains the expressions for the $\eta$-deformed Neumann model studied in  \cite{Arutyunov:2014cda}. Moreover, one can also consider the limit $\omega_{i}\rightarrow0$ in \eqref{solucion2}, which would correspond to solutions depending exclusively on the world-sheet coordinate $\sigma$. Then, by performing the double Wick rotation $\sigma\leftrightarrow\tau$ (which leaves the Lagrangian \eqref{FulletaLag} invariant up to an overall minus), we see that the corresponding classical solution describes geodesic motion on the $\eta$-deformed sphere. As was explained in \cite{Arutyunov:2014cda}, geodesics on this background are described by an integrable deformation of the Rosochatius system. Thus, by studying the $\eta$-deformed 
Neumann-Rosochatius model and its relevant limits, one can also obtain integrals for geodesic motion on the deformed background.

For a diagram describing the relation between these three different classical solutions and their corresponding deformed and undeformed integrable models see Figure \ref{noche12}.


\section{Lax pair for the $\eta$-deformed Neumann-Rosochatius model}\label{construyendo}

We will proceed to construct a $4\times4$ Lax representation for the system presented in the previous section, and later on, we will use it to create the $\eta$-deformed analogues of the integrals of motion \eqref{IntNR}. The procedure is in spirit similar to the one used in \cite{Arutyunov:2014cda}, although expressions are considerably more complicated. Therefore, in our exposition we will briefly introduce the key points in this construction, omitting intermediate expressions in the derivations.

\subsection{Construction of a $4\times4$ Lax representation}

Our starting point is the zero-curvature representation for the bosonic sigma model of $(\ads)_{\eta}$, as proposed in \cite{Delduc:2013qra},
\begin{align}
 \partial_{\alpha}\mathcal{L}_{\beta}-\partial_{\beta}\mathcal{L}_{\alpha}+\left[\mathcal{L}_{\alpha},\mathcal{L}_{\beta}\right]=0\ ,
\label{zerocurvatureetadeformations}
\end{align}
where the Lax connection $\mathcal{L}_{\alpha}$ consists of $8\times8$ matrices defined by
\begin{align*}
\mathcal{L}^{\alpha}&=\widetilde{J}_{+}^{\alpha (0)}+J_{-}^{\alpha (0)}+\lambda^{-1}\sqrt{1+\varkappa^{2}}\ \widetilde{J}_{+}^{\alpha (2)}+\lambda\sqrt{1+\varkappa^{2}}\ J_{-}^{\alpha (2)}\ .
\end{align*}
Here $\lambda$ is the spectral parameter, $J^{\alpha(i)}$ and $\widetilde{J}^{\alpha(i)}$ denote the respective components of $J^{\alpha}$ and $\widetilde{J}^{\alpha}$ along the $\mathbb{Z}_{4}$ graded subspaces $i=0,...,3$, while $J_ - ^\alpha$ and $\widetilde J_ + ^\alpha$  are projections  $J_{-}^{\alpha}=P^{\alpha\beta}_{-}J_{\beta}$ and $\widetilde{J}_{+}^{\alpha}=P^{\alpha\beta}_{+}\widetilde{J}_{\beta}$, with $P_{\pm}^{\alpha\beta}=\frac{1}{2}(\eta^{\alpha\beta}\pm\epsilon^{\alpha\beta})$, of the deformed currents
\begin{align}
J_{\alpha } = -\frac{1}{1 - \varkappa R_{\mathfrak{g}} \circ P_{2}}\left(A_{\alpha} \right),\ \ \ \ \ \ \ \ \ \ \ \ \ \ \widetilde{J}_{\alpha } = -\frac{1}{1 + \varkappa R_{\mathfrak{g}} \circ P_{2}}\left(A_{\alpha} \right).
\label{definitionJs}
\end{align}
In the expressions above $P_{i}$ is the projector along the subspace with grade $i=0,..,3$ and we used the definition $A_{\alpha}=-\mathfrak{g}^{-1}\partial_{\alpha}\mathfrak{g}$, where $\mathfrak{g}=\mathfrak{g}(\tau,\sigma)$ denotes a bosonic coset representative of ${\rm SU}(2,2)\times {\rm SU}(4)/{\rm SO}(4,1) \times {\rm SO}(5)$. Additionally, for the operator $R_{\mathfrak{g}}$ on $M\in \mathfrak{psu}(2,2|4)$ we use the definition
$$R_{\mathfrak{g}}(M)=\mathfrak{g}^{-1}R(\mathfrak{g}M\mathfrak{g}^{-1})\mathfrak{g} ,$$
where $R$ is an operator on $\mathfrak{psu}(2,2|4)$ satisfying the modified Yang-Baxter equation \cite{Delduc:2013qra}. By choosing the same coset representative $\mathfrak{g}(\tau,\sigma)$ and $R$, as in \cite{Arutyunov:2013ega},
$$R(M)_{ij}=-i\epsilon_{ij} M_{ij}, \ \ \ \ \  \epsilon_{ij}= \left\{
	\begin{array}{ll}
		\ \ 1\ \ \text{if}\  i<j \\
		\ \ 0\ \ \text{if} \ i=j\\
 		-1\ \ \text{if} \ i>j
	\end{array}
\right. ,$$
and inverting the operators $1 \pm \varkappa R_{\mathfrak{g}}\circ P_{2}$ in \eqref{definitionJs}, one obtains an explicit expression  for the $8\times8$ Lax connection $\mathcal{L}_{\alpha}$.

Plugging in the generalised spinning solution \eqref{solucion2}, the coordinate $\tau$ decouples, as the coordinates $r$, $\xi$ and $\alpha_{i}$ depend exclusively on $\sigma$, and the zero-curvature condition \eqref{zerocurvatureetadeformations} reduces to
\begin{align}
\partial_{\sigma}\mathcal{L}_{\tau}=\left[\mathcal{L}_{\tau},\mathcal{L}_{\sigma}\right] .
\label{zerocurvaturespinning}
\end{align}
In the bosonic sigma model, the $8\times8$ Lax connection takes a block-diagonal form, with one $4\times4$ block corresponding to $\rm{AdS}_{5}$  and the other one corresponding to $\rm{S}^{5}$. As our classical solution lives on $\rm{S}^{5}$, we restrict ourselves to the latter. 

By explicit calculation, it can be checked that equation \eqref{zerocurvaturespinning} is satisfied after imposing the equations of motion coming from \eqref{LagNR} and therefore, \eqref{zerocurvaturespinning} is a $4\times4$ Lax pair for our system. Naturally, this Lax pair reduces to the one constructed for spinning solutions in \cite{Arutyunov:2014cda} when taking the limit $\alpha_{i}\rightarrow0$ and $\pi_{\alpha_{i}}\rightarrow0$, and it is related to the Lax pair used to study geodesics in \cite{Arutyunov:2014cda} after taking the limit $\omega_{i}\rightarrow0$ and considering the time parameter of the system to be $\tau$ instead of $\sigma$. 

The entries of this pair of $4\times4$ matrices constitute very large expressions which we present in Appendix \ref{laxapendex}. Since we are mainly interested in the deformed analogues of the integrals \eqref{IntNR}, we will now proceed to use $\mathcal{L}_{\tau}$ to generate a tower of integrals of motion.

\subsection{Generating deformed integrals of motion $\widetilde{I}_{i}$}\label{findingI}

Conserved quantities for this system can be generated by considering the trace of powers of $\mathcal{L}_{\tau}$, namely $\text{Tr}\left[ {\mathcal{L}_\tau ^n}(\lambda) \right]$, and then series expand in the spectral parameter $\lambda$. Due to Newton's identities for the traces, it suffices to restrict our analysis to $n\leq4$.

By explicit calculation it can be checked that for $n\leq3$, all integrals of motion generated in this way  can be written in terms of $\pi_{\alpha_{i}}$ and $\widetilde{H}_{NR}$. However, for the case of $n=4$, the integrals of motion generated by
 $$K_{m}={\left. {\frac{1}{{m!}}\frac{{{d^m}\left( {{\lambda ^4}\mathcal{L}_\tau ^4} \right)}}{{d{\lambda ^m}}}} \right|_{\lambda  = 0}}\quad\quad \text{with}\quad  m \in \left\{ {2,4,6} \right\} ,$$
  can be decomposed in terms of the previously known integrals and a new unknown integral of motion, which will be denoted by $\widetilde{K}$.
 
This new integral of motion $\widetilde{K}$ comes as a very large expression, having terms up to quadratic order in $\pi_{r}$ and quartic order in $\pi_{\xi}$. In principle, this integral together with the momenta $\pi_{\alpha_{i}}$ and $\widetilde{H}_{NR}$, constitute an involutive family, ensuring classical integrability in the Liouville sense. However, the undeformed limit of this quantity suggests that it can be split further into smaller conserved blocks, namely into integrals $\widetilde{I}_{i}$ deforming the ones in equation \eqref{IntNR}. 

As in the undeformed case, it is expected that only two of the deformed integrals $\widetilde{I}_{i}$ are truly independent, see \eqref{propertiesundefNR2}, and that the Hamiltonian is given by a linear combination of the three $\widetilde{I}_{i}$, see \eqref{propertiesundefNR3}. Thus, we adopt the following ansatz
\begin{align*}
{{\widetilde H}_{NR}} &= {A_1}{{\widetilde I}_1} + {A_2}{{\widetilde I}_2} + {A_3}\ ,\\
\widetilde K &= {B_1}{{\widetilde I}_1} + {B_2}{{\widetilde I}_2} + {B_3}\ ,
\end{align*}
where $A_{i}$ and $B_{i}$ are constants independent of $\sigma$, which in principle can depend on $\pi_{\alpha_{i}}$, $\omega_{i}$ and $\varkappa$. By using the explicit expressions of $\widetilde{K}$ and $\widetilde{H}_{NR}$, one can solve for $\widetilde{I}_{1}$ and $\widetilde{I}_{2}$ in terms of the $A_{i}$ and $B_{i}$.

From the expressions \eqref{HNR} and \eqref{IntNR}, we see that the undeformed system satisfies
$$\frac{{{\partial ^2}{H_{NR}}}}{{\partial \omega _i^2}} = {\left. {{{I}_i}} \right|_{{\pi _r}=\pi_{\xi} = {\pi _{{\alpha _j}}} = 0}}\quad \forall i \in \left\{ {1,2,3} \right\}.$$
From equation \eqref{HamNrrr} one also has that ${\partial ^2}{{\widetilde H}_{NR}}/\partial \omega _i^2 = {\partial ^2}{H_{NR}}/\partial \omega _i^2$. Since we want the $\eta$-deformed integrals $\widetilde{I}_{i}$ to coincide with the undeformed ones in the limit of $\varkappa\rightarrow0$, we will impose that
\begin{align}\label{thetrixk}
\frac{{{\partial ^2}{{\widetilde H}_{NR}}}}{{\partial \omega _i^2}} = {\left. {{{\widetilde I}_i}} \right|_{{\pi _r}=\pi_{\xi} = {\pi _{{\alpha _j}}} = 0}}\quad \forall i \in \left\{ {1,2} \right\},
\end{align}
which is in essence equivalent to imposing $${\left. {{I_i}} \right|_{{\pi _r}=\pi_{\xi} = {\pi _{{\alpha _j}}} = 0}} = {\left. {{{\widetilde I}_i}} \right|_{{\pi _r}=\pi_{\xi} = {\pi _{{\alpha _j}}} = 0}}\ .$$ 

By replacing the $\widetilde{I}_{i}(A_{i},B_{i})$ on the right hand side of \eqref{thetrixk}, we find conditions for the constant coefficients $A_{i}$ and $B_{i}$. For instance, one can start with $i=1$, where one needs to match both sides of the equation \eqref{thetrixk} term by term in powers of $r(\sigma)$. By doings this, one finds the coefficients $B_{i}$ in terms of the $A_{i}$. Repeating this procedure for the case of $i=2$, equation \eqref{thetrixk} fixes the constants $A_{i}$ after matching both sides in powers of $r(\sigma)$. In analogy to \eqref{propertiesundefNR2}, we define the integral $\widetilde{I}_{3}$ through the relation
\begin{align}\label{ela0}
1 = {{\widetilde I}_1} + {{\widetilde I}_2} + {{\widetilde I}_3}\ .
\end{align}
By explicit calculation, one can check that the deformed integrals of motion obtained in this way satisfy
\begin{align}\label{ela1}
{{\widetilde H}_{NR}} = \frac{1}{2}\sum\limits_{i = 1}^3 {\left( {\omega _i^2{{\widetilde I}_i} + \pi _{{\alpha _i}}^2} \right)} \ ,
\end{align}
and
\begin{align}\label{ela2}
\left\{ {{{\widetilde H}_{NR}},{{\widetilde I}_i}} \right\}_{P.B.} = 0\ ,&&\left\{ {{{\widetilde I}_i},{{\widetilde I}_j}} \right\}_{P.B.} = 0\ ,&& \left\{ {{{\widetilde I}_i},{\pi _{{\alpha _j}}}} \right\}_{P.B.} = 0\ ,
\end{align}
where we used the standard Poisson brackets of the unconstrained $(r,\xi)$ coordinates.

By looking at the integrals $\widetilde{I}_{i}$ obtained by this procedure term by term, one sees that they contain a few constant terms proportional to powers of the momenta $\pi_{\alpha_{i}}$. These terms can be removed such that \eqref{ela0}, \eqref{ela1} and \eqref{ela2} are left unchanged. The final results for the $\eta$-deformed integrals of motion obtained by this construction are presented in equations \eqref{IntDefNR1}, \eqref{IntDefNR2} and \eqref{IntDefNR3}.

\subsection{Moving from $(r,\xi)$ to $x_{i}$ coordinates}

So far, all our calculations have been performed in the unconstrained coordinates $(r,\xi)$. However, it is convenient to transform the Hamiltonian and integrals of motion to the constrained coordinates $x_{i}$ described in  equation \eqref{cocord}, since these are the ones commonly used in the literature. The map to the $x_{i}$ coordinates used here was first derived in \cite{Arutyunov:2014cda} and consists of several steps, which we will briefly explain.

First, we write the deformed integrals of motion $\widetilde{I}_{i}$ in terms of $r'\left( \sigma  \right)$ and $\xi '\left( \sigma  \right)$ instead of momenta $\pi_{r}$ and $\pi_{\xi}$; this is easily achieved by making use of ${\pi _r} = \partial {{\widetilde L}_{NR}}/\partial r'$,  ${\pi _\xi} = \partial {{\widetilde L}_{NR}}/\partial \xi'$ and the Lagrangian \eqref{LagNR}. Then we proceed to perform a change of coordinates $\{r,\xi \}\rightarrow\{x_{1},x_{2}\}$ by means of
\begin{align*}
r = \sqrt {x_1^2 + x_2^2} \ ,&& \xi  = \arctan \left( {\frac{{{x_2}}}{{{x_1}}}} \right)\ .
\end{align*}
Once the Lagrangian and deformed integrals of motion are written in terms of $x_{1}$, $x_{2}$ and their derivatives, we calculate the momenta ${p_1} = \partial {{\widetilde L}_{NR}}/\partial {x'_1}$ and ${p_2} = \partial {{\widetilde L}_{NR}}/\partial {x'_2}$ conjugated to $x_{1}$ and $x_{2}$, respectively. Using these results, we proceed to rewrite the deformed integrals in terms of phase-space variables such that they have the functional dependence $\widetilde{I}_{i}(x_{1},x_{2},p_{1},p_{2})$. Afterwards, a third coordinate is introduced through the relation
\begin{align}\label{constrait1a}
1 = x_1^2 + x_2^2 + x_3^2\ .
\end{align}
In order to introduce a momentum conjugate to $x_{3}$, we perform the transformation
\begin{align*}
{p_1} \to \frac{{{\pi _1}{x_3} - {\pi _3}{x_1}}}{{{x_3}}}\ ,&& {p_2} \to \frac{{{\pi _2}{x_3} - {\pi _3}{x_2}}}{{{x_3}}}\ ,
\end{align*}
where we have yet to express the new momenta $\pi_{i}$ with $i\in\{1,2,3\}$ in terms of $p_{1}$, $p_{2}$ and the $x_{i}$.

By performing this procedure to the integrals of motion found in subsection \ref{findingI}, we find the expressions for the $\widetilde{I}_{i}$ in terms of the phase-space coordinates $x_{i}$ and $\pi_{i}$. Writing the Hamiltonian in these coordinates is easily achieved by the use of $\eqref{ela1}$, where we defer presentation of the resulting expressions for $\widetilde{H}_{NR}$ and $\widetilde{I}_{i}$ to Section \ref{Defsystems}.

As seen before, the undeformed system is subjected to both the constraints of equation \eqref{undefconst}. In the $x_{i}$ coordinates, the deformed model immediately satisfies the first constraint \eqref{constrait1a}, and we would like to impose also for the new momenta $\pi_{i}$ to satisfy the second constraint
\begin{align}\label{constrait1b}
\sum\limits_{i = 1}^3 {{\pi _i}{x_i} = 0} \ .
\end{align}
In order to find an explicit expression for the $\pi_{i}$ in terms of $x_{i}$ and $x'_{i}$, we express the Hamiltonian $\widetilde{H}_{NR}$ in terms of the new phase-space coordinates $x_{i}$ and $\pi_{i}$, and then use the equations
\begin{align}\label{piconditions}
{x'_1} = \left\{ {{{\widetilde H}_{NR}},{x_1}} \right\}_{D.B.}\ , && {x'_2} = \left\{ {{{\widetilde H}_{NR}},{x_2}} \right\}_{D.B.}\ ,
\end{align}
where Dirac brackets are required due to the constraints \eqref{constrait1a} and \eqref{constrait1b}. By explicit evaluation of the right hand sides in \eqref{piconditions}, we end up with two extra equations relating the momenta $\pi_{i}$ with the $x_{i}$ and $x'_{i}$. By solving for $\pi_{i}$ in equations \eqref{constrait1b} and \eqref{piconditions}, one obtains
\begin{align}
\pi_{1}=&\frac{1}{u}\left[{x'_1} - \varkappa {x_1}\left( {x_2^2{\omega _1} + x_3^2{\omega _3}} \right) + {\varkappa ^2}{x_2}\left( {{x_2}{x'_1}\left( {1 + x_1^2} \right) - {x_1}{x'_2}\left( {1 - x_2^2} \right)} \right) \right. \nonumber\\
& \left.	- {\varkappa ^3}{x_1}x_2^2\left( {x_1^2 + x_2^2} \right)\left( {{\omega _1} + x_3^2{\omega _3}} \right)\right],\nonumber\\
\pi_{2}=&\frac{1}{u}\left[{x'_2} - \varkappa {x_2}\left( {x_3^2{\omega _3} - x_1^2{\omega _1}} \right) + {\varkappa ^2}\left( {\left( {x_1^2 + x_2^4} \right){x'_2} - {x_1}{x_2}\left( {1 - x_2^2} \right){x'_1}} \right)\right.\label{qwerty}\\
& \left. + {\varkappa ^3}{x_2}\left( {x_1^2 + x_2^2} \right)\left( {x_1^2{\omega _1} - x_3^2x_2^2{\omega _3}} \right)\right] ,\nonumber\\
\pi_{3}=&\frac{{{x'_3} + \varkappa {\omega _3}{x_3}\left( {x_1^2 + x_2^2} \right)}}{{1 + {\varkappa ^2}\left( {x_1^2 + x_2^2} \right)}}\ ,\nonumber
\end{align}
where $u$ is given by
\begin{align*}
u = \left( {1 + {\varkappa ^2}\left( {x_1^2 + x_2^2} \right)} \right)\left( {1 + {\varkappa ^2}x_2^2\left( {x_1^2 + x_2^2} \right)} \right)\ .
\end{align*}
Naturally, in the $\varkappa\rightarrow0$ limit we see that ${\pi _i} \to {x'_i}$, as expected for the undeformed model.

As done in \cite{Arutyunov:2014cda}, a consistency check on the definition of the $\pi_{i}$ is to check that the Dirac brackets correctly determine the time evolution of the system
\begin{align*}
{x'_i} = {\left\{ {{{\widetilde H}_{NR}},{x_i}} \right\}_{D.B.}}\ ,&& {\pi'_i} = {\left\{ {{{\widetilde H}_{NR}},{\pi _i}} \right\}_{D.B.}}\ .
\end{align*}

The equation on the left holds by construction for the cases of $i=1,2$. For $i=3$, after evaluating the right hand side, one uses the explicit expressions \eqref{qwerty} for the $\pi_{i}$, the constraint \eqref{constrait1a} and its derivative. Similarly, the equation on the right can be checked by evoking \eqref{qwerty}, rewriting $x_{3}$ and $x'_{3}$ in terms of $x_{1}$, $x_{2}$ and their derivatives, and using the Euler-Lagrange equations coming from $\widetilde{L}_{NR}$ written in terms of $x_{1}$ and $x_{2}$.

For completeness, we also present the angular momenta ${\pi _{{\alpha _i}}}$ written in the $x_{i}$ coordinates
\begin{align*}
{\pi _{{\alpha _1}}} = \frac{{x_1^2{\alpha'_1}}}{{1 + {\varkappa ^2}x_2^2\left( {x_1^2 + x_2^2} \right)}}\ ,\quad\quad {\pi _{{\alpha _2}}} = x_2^2{\alpha'_2}\ ,\quad\quad {\pi _{{\alpha _3}}} = \frac{{x_3^2{\alpha'_3}}}{{1 + {\varkappa ^2}\left( {x_1^2 + x_2^2} \right)}}\ .
\end{align*}


\section{The $\eta$-deformed Neumann-Rosochatius model}\label{Defsystems}

Here we present the main results of our calculation, which are the deformed integrals of motion $\widetilde{I}_{i}$ along with the Hamiltonian $\widetilde{H}_{NR}$ in the $x_{i}$ coordinates. First, the explicit expressions for these quantities are given. Then, we elaborate on the different limits of the $\eta$-deformed Neumann-Rosochatius system by relating its integrals of motion to the ones of the $\eta$-deformed Neumann and Rosochatius models (recall Figure \ref{noche12}). Finally, we briefly discuss truncations of the $\eta$-deformed Neumann-Rosochatius to lower dimensions.

\subsection{Identities and integrals of motion}\label{mainresultsI}

The Hamiltonian for the $\eta$-deformed Neumann-Rosochatius model is given by
\begin{align}
{\widetilde H_{NR}} =& {H_{NR}} - \varkappa \left( {{\omega _1}{x_1}{x_2}{J_{12}} + {\omega _3}{x_1}{x_3}{J_{13}} + {\omega _3}{x_2}{x_3}{J_{23}}} \right)\label{HNRtildex}\\
 &+ \frac{{{\varkappa ^2}}}{2}\left[ {\left( {x_2^2 - x_3^2} \right)J_{12}^2 + \left( {x_1^2 + x_2^2} \right)J_{13}^2 + \left( {x_1^2 + x_2^2} \right)J_{23}^2} \right]\nonumber\\
  &+ \frac{{{\varkappa ^2}\left( {x_1^2 + x_2^2} \right)}}{2}\left( {\frac{{\pi _{{\alpha _1}}^2x_2^2}}{{x_1^2}} + \frac{{\pi _{{\alpha _3}}^2}}{{x_3^2}}} \right)\ ,\nonumber
\end{align}
where the undeformed Hamiltonian $H_{NR}$ is written in equation \eqref{HNR}. This system is subjected to the constraints of equations \eqref{constrait1a} and \eqref{constrait1b}.

It is interesting to point out that the first two lines in the expression for $\widetilde{H}_{NR}$ contain deformations of the Neumann potential and the kinetic term, while the third line is proportional to $\pi _{{\alpha _1}}^2$ and $\pi _{{\alpha _3}}^2$, and consequently has its origins in deformations of the Rosochatius potential. Moreover, equation \eqref{HNRtildex} exposes the asymmetry of the deformation along the different $x_i$ directions.

Liouville integrability of this model is due to the existence of deformed integrals of motion $\widetilde{I}_{i}$ derived in Section \ref{construyendo}
\begin{align}\label{I1NReta}
{\widetilde I_1} =& {I_1} + \left[ \frac{{\sum\limits_{i = 1}^4 {{n_i}{\varkappa ^i}} }}{{\left( {\omega _1^2 - \omega _2^2} \right)\left( {\omega _1^2 - \omega _3^2} \right)}} - \frac{{2\varkappa {J_{13}}{x_1}{x_3}{\omega _3}}}{{\omega _1^2 - \omega _3^2}} - \frac{{2\varkappa {J_{12}}{x_1}{x_2}{\omega _1}}}{{\omega _1^2 - \omega _2^2}}\right.\\
&\left.+ \frac{{{\varkappa ^2}J_{12}^2x_2^2}}{{\omega _1^2 - \omega _2^2}} + \frac{{{\varkappa ^2}J_{13}^2\left( {x_1^2 + x_2^2} \right)}}{{\omega _1^2 - \omega _3^2}} - \frac{{{\varkappa ^2}J_{12}^2x_3^2\omega _1^2}}{{\left( {\omega _1^2 - \omega _2^2} \right)\left( {\omega _1^2 - \omega _3^2} \right)}} \right]\nonumber\\
&+ \left[ \frac{{\varkappa \left( {\pi _{{\alpha _1}}^4{m_1} + \pi _{{\alpha _1}}^2\pi _{{\alpha _2}}^2{m_2} + \pi _{{\alpha _1}}^2\pi _{{\alpha _3}}^2{m_3} + \pi _{{\alpha _2}}^2\pi _{{\alpha _3}}^2{m_4} + \pi _{{\alpha _1}}^2{m_5} + \pi _{{\alpha _2}}^2{m_6} + \pi _{{\alpha _3}}^2{m_7}} \right)}}{{\left( {\omega _1^2 - \omega _2^2} \right)\left( {\omega _1^2 - \omega _3^2} \right)}}\right.\nonumber\\
&\left.+ \frac{{{\varkappa ^2}\omega _1^2\pi _{{\alpha _1}}^2x_2^2\left( {1 - x_3^2} \right)}}{{\left( {\omega _1^2 - \omega _2^2} \right)\left( {\omega _1^2 - \omega _3^2} \right)x_1^2}} + \frac{{{\varkappa ^2}\omega _1^2\pi _{{\alpha _3}}^2\left( {1 - x_3^2} \right)}}{{\left( {\omega _1^2 - \omega _2^2} \right)\left( {\omega _1^2 - \omega _3^2} \right)x_3^2}} \right]\ ,\nonumber
\end{align}
\begin{align}\label{I2NReta}
{\widetilde I_2} =& {I_2} + \left[ \frac{{\sum\limits_{i = 1}^4 {{n_i}{\varkappa ^i}} }}{{\left( {\omega _2^2 - \omega _1^2} \right)\left( {\omega _2^2 - \omega _3^2} \right)}} - \frac{{2\varkappa {J_{23}}{x_2}{x_3}{\omega _3}}}{{\omega _2^2 - \omega _3^2}} - \frac{{2\varkappa {J_{12}}{x_1}{x_2}{\omega _1}}}{{\omega _2^2 - \omega _1^2}}\right.\\
 &\left.+ \frac{{{\varkappa ^2}J_{12}^2x_2^2}}{{\omega _2^2 - \omega _1^2}} + \frac{{{\varkappa ^2}J_{23}^2\left( {x_1^2 + x_2^2} \right)}}{{\omega _2^2 - \omega _3^2}} - \frac{{{\varkappa ^2}J_{12}^2x_3^2\omega _2^2}}{{\left( {\omega _2^2 - \omega _1^2} \right)\left( {\omega _2^2 - \omega _3^2} \right)}} \right]\nonumber\\
  &+ \left[ \frac{{\varkappa \left( {\pi _{{\alpha _1}}^4{m_1} + \pi _{{\alpha _1}}^2\pi _{{\alpha _2}}^2{m_2} + \pi _{{\alpha _1}}^2\pi _{{\alpha _3}}^2{m_3} + \pi _{{\alpha _2}}^2\pi _{{\alpha _3}}^2{m_4} + \pi _{{\alpha _1}}^2{m_5} + \pi _{{\alpha _2}}^2{m_6} + \pi _{{\alpha _3}}^2{m_7}} \right)}}{{\left( {\omega _2^2 - \omega _1^2} \right)\left( {\omega _2^2 - \omega _3^2} \right)}}\right.\nonumber \\
 &\left. + \frac{{{\varkappa ^2}\omega _2^2\pi _{{\alpha _1}}^2x_2^2\left( {1 - x_3^2} \right)}}{{\left( {\omega _2^2 - \omega _1^2} \right)\left( {\omega _2^2 - \omega _3^2} \right)x_1^2}} + \frac{{{\varkappa ^2}\omega _2^2\pi _{{\alpha _3}}^2\left( {1 - x_3^2} \right)}}{{\left( {\omega _2^2 - \omega _1^2} \right)\left( {\omega _2^2 - \omega _3^2} \right)x_3^2}} \right]\nonumber\ ,\nonumber
\end{align}
\begin{align}\label{I3NReta}
{\widetilde I_3} =& {I_3} + \left[ \frac{{\sum\limits_{i = 1}^4 {{n_i}{\varkappa ^i}} }}{{\left( {\omega _3^2 - \omega _1^2} \right)\left( {\omega _3^2 - \omega _2^2} \right)}} - \frac{{2\varkappa {J_{23}}{x_2}{x_3}{\omega _3}}}{{\omega _3^2 - \omega _2^2}} - \frac{{2\varkappa {J_{13}}{x_1}{x_3}{\omega _3}}}{{\omega _3^2 - \omega _1^2}}\right.\\
&\left. + \frac{{{\varkappa ^2}J_{13}^2\left( {x_1^2 + x_2^2} \right)}}{{\omega _3^2 - \omega _1^2}} + \frac{{{\varkappa ^2}J_{23}^2\left( {x_1^2 + x_2^2} \right)}}{{\omega _3^2 - \omega _2^2}} - \frac{{{\varkappa ^2}J_{12}^2x_3^2\omega _3^2}}{{\left( {\omega _3^2 - \omega _1^2} \right)\left( {\omega _3^2 - \omega _2^2} \right)}} \right]\nonumber\\
&+ \left[ \frac{{\varkappa \left( {\pi _{{\alpha _1}}^4{m_1} + \pi _{{\alpha _1}}^2\pi _{{\alpha _2}}^2{m_2} + \pi _{{\alpha _1}}^2\pi _{{\alpha _3}}^2{m_3} + \pi _{{\alpha _2}}^2\pi _{{\alpha _3}}^2{m_4} + \pi _{{\alpha _1}}^2{m_5} + \pi _{{\alpha _2}}^2{m_6} + \pi _{{\alpha _3}}^2{m_7}} \right)}}{{\left( {\omega _3^2 - \omega _1^2} \right)\left( {\omega _3^2 - \omega _2^2} \right)}}\right.\nonumber\\
&\left. + \frac{{{\varkappa ^2}\omega _3^2\pi _{{\alpha _1}}^2x_2^2\left( {1 - x_3^2} \right)}}{{\left( {\omega _3^2 - \omega _1^2} \right)\left( {\omega _3^2 - \omega _2^2} \right)x_1^2}} + \frac{{{\varkappa ^2}\omega _3^2\pi _{{\alpha _3}}^2\left( {1 - x_3^2} \right)}}{{\left( {\omega _3^2 - \omega _1^2} \right)\left( {\omega _3^2 - \omega _2^2} \right)x_3^2}} \right]\nonumber\ ,\nonumber
\end{align}
where $I_{i}$ denotes the integrals of motion of the undeformed Neumann-Rosochatius model  \eqref{IntNR}. In the expressions above, the $n_i$ are given by
\begin{align*}
n_{1}=& - 2{J_{12}}{J_{13}}{J_{23}}{\omega _1}\ ,\\
n_{2}=&- J_{12}^2x_3^2\omega _3^2 + 2{J_{12}}{x_3}{\omega _1}{\omega _3}\left( {{J_{23}}{x_1} + {J_{13}}{x_2}} \right) - J_{12}^2J_{13}^2\ ,\\
n_{3}=&2{J_{12}}{J_{13}}\left[ {{J_{12}}{x_1}{x_3}{\omega _3} - {J_{23}}\left( {x_1^2 + x_2^2} \right){\omega _1}} \right]\ ,\\
n_{4}=& - J_{12}^2J_{13}^2\left( {x_1^2 + x_2^2} \right)\ ,
\end{align*}
while the $m_i$ are defined as
\begin{align*}
{m_1} =& - \frac{{\varkappa {\kern 1pt} x_2^2x_3^2\left( {1 + {\varkappa ^2}\left( {x_1^2 + x_2^2} \right)} \right)}}{{x_1^4}}\ ,\\
{m_2} =& - \frac{{\varkappa \left( {1 + {\varkappa ^2}x_2^2} \right)x_3^2}}{{x_1^2}}\ ,\\
{m_3} =& - \frac{{\varkappa \left( {1 + {\varkappa ^2}} \right)x_2^2}}{{x_1^2}}\ ,\\
{m_4} =& - \frac{{\varkappa \left( {1 + {\varkappa ^2}} \right)x_1^2}}{{x_2^2x_3^2}}\ ,\\
{m_5} =& - \frac{{\varkappa x_3^2\left( {1 + {\varkappa ^2}\left( {1 - x_3^2} \right)} \right)J_{12}^2}}{{x_1^2}} - \frac{{{\varkappa} x_2^2\left( {1 + {\varkappa ^2}\left( {1 - x_3^2} \right)} \right)J_{13}^2}}{{x_1^2}} + \varkappa \left( {1 + {\varkappa ^2}\left( {1 - x_3^2} \right)} \right)J_{23}^2\nonumber\\
& + \frac{{2{\varkappa ^2}{\omega _3}x_2^2{x_3}{J_{13}}}}{{{x_1}}} - \frac{{2{J_{23}}{x_2}{x_3}\left( {{\varkappa ^2}{\omega _3}x_1^2 + {\omega _1}\left( {1 + {\varkappa ^2}\left( {1 - x_3^2} \right)} \right)} \right)}}{{x_1^2}}\nonumber\\
& - \frac{{\varkappa x_2^2}}{{x_1^2}}\left( {\omega _2^2x_3^2 + \omega _3^2 - 2{\omega _1}{\omega _3}x_3^2} \right),\\
{m_6} =& - \frac{{\varkappa J_{13}^2\left( {1 - x_3^2} \right)\left( {1 + {\varkappa ^2}\left( {1 - x_3^2} \right)} \right)}}{{x_2^2}} + \frac{{2{J_{13}}{x_1}{x_3}\left( {{\varkappa ^2}{\omega _3}\left( {1 - x_3^2} \right) + {\omega _1}\left( {1 + {\varkappa ^2}\left( {1 - x_3^2} \right)} \right)} \right)}}{{x_2^2}}\nonumber\\
& - \frac{{\varkappa {{({\omega _1} + {\omega _3})}^2}x_1^2x_3^2}}{{x_2^2}}\ ,\\
{m_7} =& - \frac{1}{{x_3^2}}\left[ {\varkappa \left( {1 + {\varkappa ^2}} \right)\left( {1 - x_2^2} \right)J_{12}^2 + 2\left( {1 + {\varkappa ^2}} \right){\omega _1}{x_1}{x_2}{J_{12}} + \varkappa \left( {\omega _1^2x_2^2 + \omega _2^2x_1^2} \right)} \right]\ .
\end{align*}

From the structure of equations \eqref{I1NReta}, \eqref{I2NReta} and \eqref{I3NReta}, we see that the deformed integrals of motion $\widetilde{I}_{i}$ are composed of three parts. The first part consists of the integrals of motion of the undeformed Neumann-Rosochatius system $I_{i}$, a second part corresponds to deformations already present in the deformed Uhlenbeck integrals of the $\eta$-deformed Neumann model \cite{Arutyunov:2014cda} (these contributions are enclosed in the first bracket $[\ ]$ from top to bottom), and a third contribution introduced by the deformations of the Rosochatius potential (these are in the second bracket $[\ ]$ from top to bottom).

It is curious to note that the deformed integrals of motion have terms with a double pole structure in the frequencies $\omega_{i}$. This feature will play an important role later on, when considering the limit $\omega_{i}\rightarrow0$. From the expressions above, we see that the deformed Hamiltonian $\widetilde{H}_{NR}$ is quadratic in momenta and has terms  up to the order $\varkappa^{2}$, while $\widetilde{I}_{i}$ are quartic in momenta and have terms up to the order $\varkappa^{4}$.

The Hamiltonian and the integrals of motion obtained for the $\eta$-deformed Neumann-Rosochatius model in the $x_{i}$ coordinates satisfy
\begin{equation}
\left\{ {{\widetilde{H}_{NR}},{\widetilde{I}_i}} \right\}_{D.B.} = 0\ ,\quad\quad \left\{ {{\widetilde{I}_i},{\widetilde{I}_j}} \right\}_{D.B.} = 0\ ,\quad \quad \left\{ {{\widetilde{I}_i},{\pi _{{\alpha _j}}}} \right\}_{D.B.} = 0\ ,
\end{equation}
\begin{equation}\label{eq433}
\sum\limits_{i = 1}^3 {{\widetilde{I}_i} = 1}\ , 
\end{equation}
\begin{equation}\label{eq434}
{\widetilde{H}_{NR}} = \frac{1}{2}\sum\limits_{i = 1}^3 {\left( {\omega _i^2{\widetilde{I}_i} +\pi_{{\alpha _i}}^2} \right)} \ ,
\end{equation}
which correspond to the deformed analogues of the identities \eqref{propertiesundefNR1}, \eqref{propertiesundefNR2} and \eqref{propertiesundefNR3} of the undeformed Neumann-Rosochatius system.

Naturally, in the undeformed limit, the Hamiltonian and the new integrals of motion reduce to the known expressions for the undeformed Neumann-Rosochatius system
\begin{align}
\mathop {\lim }\limits_{\varkappa  \to 0} {{\widetilde H}_{NR}} = {H_{NR}}\ ,&& \mathop {\lim }\limits_{\varkappa  \to 0} {{\widetilde I}_i} = {I_i}\ .
\end{align}


\subsection{Connections with the $\eta$-deformed Neumann and Rosochatius systems}\label{connectionsRandN}

As discussed in Section \ref{section3}, in the $(\ads)_{\eta}$ background, the generalised spinning solution of equation \eqref{solucion2} reduces to the spinning solution studied in \cite{Arutyunov:2014cda} by taking the limit $\alpha_{i}\rightarrow0$ and $\pi_{\alpha_{i}}\rightarrow0$, and it is connected to geodesic solutions by considering the limit $\omega_{i}\rightarrow0$ and changing the time parameter $\sigma\rightarrow\tau$. Therefore, by considering the respective limits for the $\eta$-deformed Neumann-Rosochatius model, the system must reduce to the integrable models describing spinning solutions and geodesics on $(\ads)_{\eta}$: The $\eta$-deformed Neumann and Rosochatius models, respectively (see Figure \ref{noche12}). In this section, the connection between these deformed integrable models is made explicit by considering their Hamiltonians and integrals of motion.

Reduction of the $\eta$-deformed Neumann-Rosochatius system to the $\eta$-deformed Neumann model follows straightforwardly from the results of \cite{Arutyunov:2014cda} and the expressions for $\widetilde{H}_{NR}$ and $\widetilde{I}_{i}$
\begin{align}\label{limitsystem1}
\mathop {\lim }\limits_{{\pi _{\alpha j}} \to 0} {{\widetilde H}_{NR}} = {{\widetilde H}_N}\ ,&& \mathop {\lim }\limits_{{\pi _{\alpha j}} \to 0} {{\widetilde I}_i} = {{\widetilde F}_i}\ ,
\end{align}
where $\widetilde{H}_{N}$ denotes the Hamiltonian of the $\eta$-deformed Neumann model, while $\widetilde{F}_{i}$ are the deformed Uhlenbeck integrals of the $\eta$-deformed Neumann model obtained in \cite{Arutyunov:2014cda}. Equation \eqref{limitsystem1} is therefore, the deformed counterpart of equations \eqref{counterpart1a} and  \eqref{counterpart1b}.

For the discussion on the relation between the $\eta$-deformed Neumann-Rosochatius model and geodesic solutions, we will use the expressions in $(r,\xi)$ coordinates presented in appendix \ref{definrxi}, since these are the coordinates in which integrability of geodesics was originally studied (see \cite{Arutyunov:2014cda} for details).

In Section \ref{UndefSystems}, we saw how  the integrals of motion of the undeformed Rosochatius system are obtained by taking the limit $\omega_{i}\rightarrow0$ of the expressions for the integrals of the undeformed Neumann-Rosochatius model. We will now perform a similar limit on the integrals of the $\eta$-deformed Neumann-Rosochatius system presented in equations \eqref{IntDefNR1}, \eqref{IntDefNR2} and \eqref{IntDefNR3}, with the aim of obtaining integrals of motion for the $\eta$-deformed Rosochatius model.

Due to the double pole structure that the integrals $\widetilde{I}_{i}$ have on the frequencies $\omega_{i}$, it is not sufficient to multiply by an $(\omega_{i}^{2}-\omega_{j}^{2})$ factor and then take the limit $\omega_{j}\rightarrow\omega_{i}$, as was done for the undeformed case in equation \eqref{NRtoNUndefLimit}. Instead, we will consider the following limit
\begin{equation}\label{NRtoRDeflimit}
\mathop {\lim }\limits_{{\omega _i}\to 0}\ \ \mathop {\lim }\limits_{{\omega _k},{\omega _l}\to {\omega _i} } \ {\widetilde I_i}\ \prod\limits_{j \ne i}^3 {\left( {\omega _i^2 - \omega _j^2} \right)}  =  - {\varkappa ^2}Q + {\varkappa ^4}\left( {\pi _{{\alpha _1}}^2\pi _{{\alpha _2}}^2 + \pi _{{\alpha _1}}^2\pi _{{\alpha _3}}^2 + \pi _{{\alpha _2}}^2\pi _{{\alpha _3}}^2} \right)\quad \forall i \in \left\{ {1,2,3} \right\}\ ,
\end{equation}
where there is no summation over the index $i$ on the left side and the three indices ($i$, $k$,$l$) take all different values. It can be checked explicitly that the quantity on the right hand side is indeed an integral of motion for $\eta$-deformed geodesic solutions: It is composed out of the angular momenta $\pi_{\alpha_i}$ and the integral of motion $Q$, which was obtained in \cite{Arutyunov:2014cda} through the Lax formalism. The explicit expression for $Q$ is given by
\begin{align*}
Q=&\left( {1 + {\varkappa ^2}{r^2}} \right)\left( {1 - {r^2}} \right)\bigg[ {\frac{{\pi _\xi ^4{{\sin }^2}\xi }}{{{r^2}}}}{ - \frac{{\pi _\xi ^3{\pi _r}\sin 2\xi }}{r}}{ + \pi _r^2\pi _\xi ^2{{\cos }^2}\xi }{ + \frac{{2\pi _\xi ^2\pi _{{\alpha _1}}^2{{\tan }^2}\xi }}{{{r^2}}}}\label{integralQ}\\
&{ + \frac{{\pi _\xi ^2\pi _{{\alpha_2}}^2}}{{{r^2}}}}{ + \pi _r^2\pi _{{\alpha _2}}^2{{\cot }^2}\xi }{ - \frac{{2{\pi _r}{\pi _\xi }\pi _{{\alpha _1}}^2\tan \xi }}{r}}{ - \frac{{2{\pi _r}{\pi _\xi }\pi _{{\alpha _2}}^2\cot \xi }}{r}}{ + \frac{{\pi _{{\alpha _1}}^4{{\tan }^2}\xi\  {{\sec }^2}\xi }}{{{r^2}}}}\bigg]\nonumber\\
&+ \pi _{{\alpha _1}}^2\pi _{\alpha_{3}} ^2\left( {{\varkappa ^2} + {{\sin }^2}\xi } \right){\sec ^2}\xi + \frac{{\pi _{{\alpha _2}}^2\pi _{\alpha_{3}} ^2\left( {{{\cos }^2}\xi  + {\varkappa ^2}\left( {1 - {r^2}{{\sin }^2}\xi } \right)} \right){{\csc }^2}\xi }}{{1 - {r^2}}}\nonumber\\
 &+ \frac{{\left( {1 + {\varkappa ^2}} \right)\pi _\xi ^2\pi _{\alpha_{3}} ^2\left( {1 - {r^2}{{\sin }^2}\xi } \right)}}{{1 - {r^2}}}+ \frac{{\pi _{{\alpha _1}}^2\pi _{{\alpha _2}}^2\left( {1 + {\varkappa ^2}{r^2} - {r^2}\left( {1 + {\varkappa ^2}{r^2}{{\sin }^2}\xi } \right)} \right){{\sec }^2}\xi }}{{{r^2}}}\nonumber\ .
\end{align*}
It is interesting to point out that when taking the limit $\omega_{i}\rightarrow0$, instead of the deformed analogues of the three integrals $F_{ij}$ of the undeformed Rosochatius system, for $\varkappa>0$ one obtains only one integral of motion $Q$ (independent of the $\pi_{\alpha_{i}}$). 

By considering the $\omega_{i}\rightarrow0$ limit of $\widetilde{H}_{NR}$ (recall equation \eqref{HamNrrr}), we see that
\begin{align}
\mathop {\lim }\limits_{{\omega _i} \to 0} {{\widetilde H}_{NR}} = {{\widetilde H}_R}\ ,
\end{align}
where $\widetilde{H}_{R}$ is the Hamiltonian describing geodesics in the deformed sphere, which was first calculated in \cite{Arutyunov:2014cda}. This equation, along with equation \eqref{NRtoRDeflimit}, are the deformed analogues of equations \eqref{HNRtoHR} and \eqref{NRtoNUndefLimit}, respectively.

\subsection{Lower dimensional truncations}\label{truncations}

By examining the Dirac bracket of the Hamiltonian $\widetilde{H}_{NR}$ with the coordinate $x_{i}$ and the conjugate momenta $\pi_{i}$ (with $i
\in\left\{ {1,2,3} \right\}$), it can be checked that the following is a consistent truncation of the system along the $i$-direction
\begin{align*}
x_{i}=0\ ,&& \pi_{i}=0\ ,&&\pi_{\alpha_{i}}=0\ , && \omega_{i}=0\ .
\end{align*}

In this case, the phase-space constraints of equations \eqref{constrait1a} and \eqref{constrait1b} reduce to
\begin{align*}
\sum\limits_{j \ne i}^3 {x_j^2 = 1} \ ,&& \sum\limits_{j \ne i}^3 {{x_j}{\pi _j} = 0} \ ,
\end{align*}
while the deformed integrals of motion reduce to
\begin{align*}
\mathop {\lim }\limits_{{x_i},{\pi _i},{\pi _{{\alpha _i}}},{\omega _i} \to 0} {{\widetilde I}_i} = 0\ , && \mathop {\lim }\limits_{{x_i},{\pi _i},{\pi _{{\alpha _i}}},{\omega _i} \to 0} {{\widetilde I}_j} \ne 0\ \ \ \ \ \forall j \ne i\ .
\end{align*}
The above expressions imply that equations \eqref{eq433} and \eqref{eq434} become
\begin{align*}
\sum\limits_{j \ne i}^3 {{{\left. {{{\widetilde I}_j}} \right|}_{{\pi _i} = {x_i} = {\pi _{{\alpha _i}}} = {\omega _i} = 0}} = 1} \ , && {{\widetilde H}_{NR}} = \frac{1}{2}\sum\limits_{j \ne i}^3 {{{\left. {\left( {\omega _j^2{{\widetilde I}_j} + \pi _{{\alpha _j}}^2} \right)} \right|}_{{\pi _i} = {x_i} = {\pi _{{\alpha _i}}} = {\omega _i} = 0}}} \ .
\end{align*}
Therefore, out of ${\widetilde{H}}_{NR}$ and the 2 integrals of motion $\widetilde{I}_{j}$ (with $j\neq i$), there is only one truly independent integral of motion. This is consistent with Liouville's theorem, as in the truncated system one has a 2-dimensional phase-space, once the constraints are taken into account.

Because of the asymmetry of the system along the different directions, it is possible to truncate the $\eta$-deformed Neumann-Rosochatius model in several ways. In particular, the truncation along $i=3$ corresponds to the system studied in \cite{Hernandez2016},  where their corresponding integrals of motion and identities coincide with the ones obtained by the truncation procedure explained above.


\section{Conclusions}

In the present paper we have studied the integrable model describing generalised spinning solutions in the $\eta$-deformed $\ads$ background, constituting a one-parameter deformation of the well-known Neumann-Rosochatius integrable system. By explicit construction of a $4\times4$ Lax pair representation and a set of integrals of motion in involution, we exposed the Liouville integrability of the model. The deformed integrals of motion obtained generalise the ones previously found for the Neumann model and geodesic motion on the $\eta$-deformed sphere. The construction of the integrals of motion and Lax representation for this model is a necessary first step towards finding its exact solution, however, there are still many open questions to be addressed.
 
The deformed model we considered corresponds to the $N=3$ Neumann-Rosochatius system, where motion is constrained to a two-sphere. Generally, the Neumann-Rosochatius model is known to be integrable for arbitrary $N$, where motion is constrained to a $(N-1)$-sphere. Thus, it would be interesting to generalise the results found here for $N>3$. Asymmetry of the deformation in the different $x_i$ directions makes this a very non-trivial task. A natural starting point would be to deform the sigma model on the coset space $\rm{SO}(N+1)/\rm{SO}(N)$ and then to consider a generalised spinning solution similar to the one in \eqref{solucion2}.

In Section \ref{connectionsRandN}, we explored the connection between the $\eta$-deformed Neumann-Roso-chatius and Neumann model by studying their Hamiltonians and integrals of motion. For the undeformed case integrability of the $N=3$ Neumann-Rosochatius model also follows from the fact that it can be seen as a special case of an $N=6$ Neumann model with degenerate frequencies. In fact, the undeformed integrals $I_{i}$ can be constructed explicitly by considering convenient linear combinations of the $F_{i}$ integrals of motion of the $N=6$ degenerate Neumann model. It would be interesting to see if this also holds for their $\eta$-deformed counterparts, which again would require an in depth understanding of the $N>3$ deformed models.

The $\eta$-deformed models considered in the present paper have highly complicated integrals of motion, making separation of variables a very difficult problem. A solution to this problem is supposedly given by Sklyanin's method \cite{Sklyanin:1995bm}, which yields canonical coordinates in terms of (properly normalized) eigenvalues and poles of the Baker-Akhiezer function. But because of the sheer size of the Lax pair (see Appendix \ref{laxapendex}), the corresponding equations appear to be rather involved and solutions seem difficult to find. For this reason, it would be desirable to devise a lower dimensional Lax pair.

Geodesic motion on spheres is a renowned problem, partially due to the fact that it is superintegrable. Concerning its Liouville integrability for ${\rm S}^5$, a set of integrals of motion in involution is given by the angular momenta $\pi_{{\alpha}_{i}}$, the Hamiltonian $H_{R}$ and one of the three non-abelian integrals $F_{ij}$. The fact that this system has more integrals of motion than required by Liouville's theorem imposes strong constraints on its dynamics: Geodesics are closed and the motion is periodic. For the $\eta$-deformed sphere, it was found in \cite{Arutyunov:2014cda} that geodesics are Liouville integrable due to the set of integrals of motion in involution $\pi_{{\alpha}_{i}}$, $\widetilde{H}_{R}$ and $Q$. Here, by considering a geodesic limit of the deformed Neumann-Rosochatius system, we end up with the same set of integrals of motion, leaving superintegrability an open question. An interesting way to approach this problem is to consider a lower-dimensional model obtained from geodesic motion on $({\rm S}^5)_\eta$ under the conditions $\pi_{\xi}=\pi_{\alpha_{1}}=\pi_{\alpha_{2}}=0$, such that, as was shown in \cite{Hoare:2014pna}, the corresponding motion is constrained to the manifold of Fateev's sausage model \cite{Fateev:1992tk}. Investigation of this problem is on the way \cite{Arutyunov:2016kve}.


\section*{Acknowledgements}
We would like to thank A. Dekel, J. M. Nieto, and S. C. Vargas for useful discussions and R. Klabbers for useful comments on the manuscript. M.H. thanks NORDITA for hospitality.
The work of G.A. and M.H. is supported by the German Science Foundation (DFG) under the Collaborative Research Center (SFB) 676 Particles, Strings and the Early Universe. The work of D.M. is supported by the ERC advanced grant No 341222.

\appendix
\section{$4\times4$ Lax pair for the $\eta$-deformed Neumann-Rosochatius model}\label{laxapendex}
In order to present the Lax pair for the $\eta$-deformed Neumann-Rosochatius model in a short form, we will first define
\begin{align}
{\gamma ^1} = \left( {\begin{array}{*{20}{c}}
0&0&0&{ - 1}\\
0&0&1&0\\
0&1&0&0\\
{ - 1}&0&0&0
\end{array}} \right) ,&&
{\gamma ^2} = \left( {\begin{array}{*{20}{c}}
0&0&0&i\\
0&0&i&0\\
0&{ - i}&0&0\\
{ - i}&0&0&0
\end{array}} \right) ,&&
{\gamma ^3} = \left( {\begin{array}{*{20}{c}}
0&0&1&0\\
0&0&0&1\\
1&0&0&0\\
0&1&0&0
\end{array}} \right) ,\nonumber
\end{align}
\begin{align}
{\gamma ^4} = \left( {\begin{array}{*{20}{c}}
0&0&{ - i}&0\\
0&0&0&i\\
i&0&0&0\\
0&{ - i}&0&0
\end{array}} \right) , &&
{\gamma ^5} = \left( {\begin{array}{*{20}{c}}
1&0&0&0\\
0&1&0&0\\
0&0&{ - 1}&0\\
0&0&0&{ - 1}
\end{array}} \right) .\nonumber
\end{align}
In terms of these matrices the generators of $\mathfrak{su}(4)$ can be written as $G_{j}=\frac{i}{2}\gamma^{j}$ and $G_{ij}=\frac{1}{4}[\gamma^{i},\gamma^{j}]$, with $i,j=1,...,5$. Using these generators, we have that the Lax pair is given by
\begin{align}
{\mathcal{L}_\tau } = \sum\limits_{i = 1}^5 {c_i^\tau } {G_i} + \sum\limits_{i < j}^{5} {c_{ij}^\tau {G_{ij}}} \ ,&& {\mathcal{L}_\sigma } = \sum\limits_{i = 1}^5 {c_i^\sigma } {G_i} + \sum\limits_{i < j}^{5} {c_{ij}^\sigma {G_{ij}}} \ ,\nonumber
\end{align}
where the $c_{i}^{\tau}$ and $c_{ij}^{\tau}$ are given by
\begin{align}
c_1^\tau  &= \frac{{\sqrt {1 + {\varkappa ^2}} \left( {\left( {1 - {r^2}} \right)\left( {1 - {\lambda ^2}} \right){\pi _r} + \varkappa r\left( {1 + {\lambda ^2}} \right){\pi _{{\alpha _3}}}} \right)}}{{2\lambda \sqrt {1 - {r^2}} }}\ ,\nonumber\\
c_2^\tau  &= \frac{{\sqrt {1 + {\varkappa ^2}} \left( {\left( {1 - {\lambda ^2}} \right){\pi _{{\alpha _1}}}{{\sec }^2}\xi  + {r^2}\left( {1 + {\lambda ^2}} \right)\left( { - {\omega _1} + \varkappa {\pi _\xi }\tan \xi } \right)} \right)\cos \xi }}{{2\lambda r}}\ ,\nonumber\\
c_3^\tau  &=  - \frac{{\sqrt {1 + {\varkappa ^2}} \left( {\left( { - 1 + {\lambda ^2}} \right){\pi _\xi } + \varkappa {r^2}\left( {1 + {\lambda ^2}} \right){\pi _{{\alpha _1}}}\tan \xi } \right)}}{{2\lambda r}}\ ,\nonumber\\
c_4^\tau  &= \frac{{\sqrt {1 + {\varkappa ^2}} \left( { - 2\left( { - 1 + {\lambda ^2}} \right){\pi _{{\alpha _2}}} - {r^2}\left( {1 + {\lambda ^2}} \right){\omega _2} + {r^2}\left( {1 + {\lambda ^2}} \right){\omega _2}\cos (2\xi )} \right)\csc \xi }}{{4\lambda r}}\ ,\nonumber\\
c_5^\tau  &= \frac{{\sqrt {1 + {\varkappa ^2}} \left( {\left( { - 1 + {\lambda ^2}} \right){\pi _{{\alpha _3}}} + \left( {1 - {r^2}} \right)\left( {1 + {\lambda ^2}} \right)\left( {\varkappa r{\pi _r} + {\omega _3}} \right)} \right)}}{{2\lambda \sqrt {1 - {r^2}} }}\ ,\nonumber
\end{align}
\begin{align}
c_{12}^\tau  &= \sqrt {1 - {r^2}} \left( {\varkappa {\pi _\xi }\sin \xi  - {\omega _1}\cos \xi } \right)\ ,
& c_{13}^\tau  &=  - {\pi _{{\alpha _1}}}\varkappa \sqrt {1 - {r^2}} \tan \xi\ , \nonumber\\
c_{14}^\tau  &=  - \sqrt {1 - {r^2}} {\omega _2}\sin \xi\ , 
&c_{15}^\tau  &= \varkappa {\pi _r} - r\left( {\varkappa r{\pi _r} + {\omega _3}} \right)\ ,\nonumber\\
c_{23}^\tau  &=  - \varkappa {\pi _\xi }\cos \xi  - {\omega _1}\sin \xi\ ,
& c_{24}^\tau  &=  - \varkappa {\pi _{{\alpha _2}}}\cot \xi\ ,\nonumber \\
c_{25}^\tau  &= \frac{{{\pi _{{\alpha _1}}}\varkappa \sqrt {1 - {r^2}} \sec \xi }}{r}\ ,
&c_{34}^\tau  &=  - {\omega _2}\cos \xi\ ,\nonumber \\
c_{35}^\tau  &= \frac{{\varkappa {\pi _\xi }\sqrt {1 - {r^2}} }}{r}\ ,
&c_{45}^\tau  &= \frac{{\varkappa {\pi _{{\alpha _2}}}\sqrt {1 - {r^2}} \csc \xi }}{r}\ ,\nonumber
\end{align}
while the $c_{i}^{\sigma}$ and $c_{ij}^{\sigma}$ are given by
\begin{align}
c_1^\sigma  &= \frac{{\sqrt {1 + {\varkappa ^2}} \left( {\left( { - 1 + {r^2}} \right)\left( {1 + {\lambda ^2}} \right){\pi _r} + \varkappa r\left( { - 1 + {\lambda ^2}} \right){\pi _{{\alpha _3}}}} \right)}}{{2\lambda \sqrt {1 - {r^2}} }}\ ,\nonumber\\
c_2^\sigma  &= \frac{{\sqrt {1 + {\varkappa ^2}} \left( { - \left( {1 + {\lambda ^2}} \right){\pi _{{\alpha _1}}}{{\sec }^2}\xi  + \left( { - 1 + {\lambda ^2}} \right){r^2}\left( { - {\omega _1} + \varkappa {\pi _\xi }\tan \xi } \right)} \right)\cos \xi }}{{2\lambda r}}\ ,\nonumber\\
c_3^\sigma  &=  - \frac{{\sqrt {1 + {\varkappa ^2}} \left( {\left( {1 + {\lambda ^2}} \right){\pi _\xi } + \varkappa {r^2}\left( { - 1 + {\lambda ^2}} \right){\pi _{{\alpha _1}}}\tan \xi } \right)}}{{2\lambda r}}\ ,\nonumber\\
c_4^\sigma  &= \frac{{\sqrt {1 + {\varkappa ^2}} \left( { - 2\left( {1 + {\lambda ^2}} \right){\pi _{{\alpha _2}}} + \left( { - 1 + {\lambda ^2}} \right){\omega _2}{r^2}\cos (2\xi ) + {\omega _2}{r^2}\left( {1 - {\lambda ^2}} \right)} \right)\csc \xi }}{{4\lambda r}}\ ,\nonumber\\
c_5^\sigma  &= \frac{{\sqrt {1 + {\varkappa ^2}} \left( {\left( {1 + {\lambda ^2}} \right){\pi _{{\alpha _3}}} - \left( {1 - {\lambda ^2}} \right)\left( {1 - {r^2}} \right)\left( {\varkappa r{\pi _r} + {\omega _3}} \right)} \right)}}{{2\lambda \sqrt {1 - {r^2}} }}\ , \nonumber
\end{align}
\begin{align}
c_{12}^\sigma  &=  - \frac{{{\pi _{{\alpha _1}}}\sqrt {1 - {r^2}} \sec \xi }}{{{r^2}}}\ ,
&c_{13}^\sigma  &=  - \frac{{{\pi _\xi }\sqrt {1 - {r^2}} }}{{{r^2}}}\ ,\nonumber\\
c_{14}^\sigma  &=  - \frac{{{\pi _{{\alpha _2}}}\sqrt {1 - {r^2}} \csc \xi }}{{{r^2}}}\ ,
&c_{15}^\sigma &=  - \frac{{r\left( {1 + {\varkappa ^2}} \right){\pi _{{\alpha _3}}}}}{{1 - {r^2}}}\ ,\nonumber\\
c_{23}^\sigma  &=  - \frac{{{\pi _{{\alpha _1}}}\left( {1 + {\varkappa ^2}{r^4}} \right)\sin \xi }}{{{r^2}{{\cos }^2}\xi }}\ ,
&c_{24}^\sigma  &=  - \varkappa {\omega _2}{r^2}\sin (\xi )\cos (\xi )\ ,\nonumber\\
c_{25}^\sigma  &= \varkappa r\sqrt {1 - {r^2}} \left( {{\omega _1}\cos \xi  - \varkappa {\pi _\xi }\sin \xi } \right)\ ,&
c_{34}^\sigma  &=  - \frac{{{\pi _{{\alpha _2}}}\cos \xi }}{{{r^2}{{\sin }^2}\xi }}\ ,\nonumber\\
c_{35}^\sigma  &= {\varkappa ^2}{\pi _{{\alpha _1}}}r\sqrt {1 - {r^2}} \tan \xi \ ,
&c_{45}^\sigma  &= \varkappa {\omega _2}r\sqrt {1 - {r^2}} \sin \xi \ .\nonumber
\end{align}

\section{The Neumann-Rosochatius model in $(r,\xi)$ coordinates}\label{undefinrxi}

The Hamiltonian for this system in $(r,\xi)$ coordinates can easily be obtained from \eqref{HamNrrr} by taking the $\varkappa\rightarrow0$. For completeness we will give the  integrals of motion $I_{i}$ in these unconstrained coordinates
\begin{align}
{I_1} =& {r^2}{\cos ^2}\xi  + \frac{1}{{\omega _1^2 - \omega _2^2}}\left[ {\pi _\xi ^2 + \pi _{{\alpha _1}}^2{{\tan }^2}\xi  + \pi _{{\alpha _2}}^2{{\cot }^2}\xi } \right] + \frac{1}{{\omega _1^2 - \omega _3^2}}\left[ \left( {1 - {r^2}} \right)\pi _r^2{{\cos }^2}\xi\right.\label{IntUndefNR1}\\
&\left. + \frac{{\pi _\xi ^2\left( {1 - {r^2}} \right){{\sin }^2}\xi }}{{{r^2}}} - \frac{{{\pi _r}{\pi _\xi }\left( {1 - {r^2}} \right)\sin 2\xi }}{r} + \frac{{\pi _{{\alpha _1}}^2\left( {1 - {r^2}} \right){{\sec }^2}\xi }}{{{r^2}}} + \frac{{\pi _{{\alpha _3}}^2{r^2}{{\cos }^2}\xi }}{{1 - {r^2}}} \right]\ ,\nonumber\\
{I_2} =& {r^2}{\sin ^2}\xi  + \frac{1}{{\omega _2^2 - \omega _1^2}}\left[ {\pi _\xi ^2 + \pi _{{\alpha _1}}^2{{\tan }^2}\xi  + \pi _{{\alpha _2}}^2{{\cot }^2}\xi } \right] + \frac{1}{{\omega _2^2 - \omega _3^2}}\left[ \pi _r^2\left( {1 - {r^2}} \right){{\sin }^2}\xi\right.\label{IntUndefNR2}\\ 
&\left. + \frac{{\pi _\xi ^2\left( {1 - {r^2}} \right){{\cos }^2}\xi }}{{{r^2}}} + \frac{{{\pi _r}{\pi _\xi }\left( {1 - {r^2}} \right)\sin 2\xi }}{r} + \frac{{\pi _{{\alpha _2}}^2\left( {1 - {r^2}} \right){{\csc }^2}\xi }}{{{r^2}}} + \frac{{\pi _{{\alpha _3}}^2{r^2}{{\sin }^2}\xi }}{{1 - {r^2}}} \right]\ ,\nonumber\\
{I_3} =& 1 - {r^2} + \frac{1}{{\omega _3^2 - \omega _1^2}}\left[ \left( {1 - {r^2}} \right)\pi _{r}^2{{\cos }^2}\xi  + \frac{{\pi _\xi ^2\left( {1 - {r^2}} \right){{\sin }^2}\xi }}{{{r^2}}}-\frac{{{\pi _r}{\pi _\xi }\left( {1 - {r^2}} \right)\sin 2\xi }}{r}\right.\label{IntUndefNR3}\\
&\left.+\frac{{\pi _{{\alpha _1}}^2\left( {1 - {r^2}} \right){{\sec }^2}\xi }}{{{r^2}}} + \frac{{\pi _{{\alpha _3}}^2{r^2}{{\cos }^2}\xi }}{{1 - {r^2}}} \right] + \frac{1}{{\omega _3^2 - \omega _2^2}}\left[ \pi _r^2\left( {1 - {r^2}} \right){{\sin }^2}\xi  \right.\nonumber\\
&\left. + \frac{{\pi _\xi ^2\left( {1 - {r^2}} \right){{\cos }^2}\xi }}{{{r^2}}} + \frac{{{\pi _r}{\pi _\xi }\left( {1 - {r^2}} \right)\sin 2\xi }}{r} + \frac{{\pi _{{\alpha _2}}^2\left( {1 - {r^2}} \right){{\csc }^2}\xi }}{{{r^2}}} + \frac{{\pi _{{\alpha _3}}^2{r^2}{{\sin }^2}\xi }}{{1 - {r^2}}}\right]\ .\nonumber
\end{align}
These expressions satisfy all the properties explained on Section \ref{UndefSystems}, but this time with Poisson brackets in the unconstrained coordinates $(r,\xi)$, instead of the Dirac brackets used for the $x_{i}$ coordinates.

\section{The $\eta$-deformed Neumann-Rosochatius model in $(r,\xi)$ coordinates}\label{definrxi}

For this integrable model, the Lagrangian and Hamiltonian in these coordinates were given in equations \eqref{LagNR} and \eqref{HamNrrr}. Here, we present the deformed integrals of motion in the $(r,\xi)$ coordinates 
\begin{align}\label{IntDefNR1}
{\widetilde I_1} =& {I_1} + \left[ \frac{{\sum\limits_{i = 1}^4 {{s_i}{\varkappa ^i}} }}{{\left( {\omega _1^2 - \omega _2^2} \right)\left( {\omega _1^2 - \omega _3^2} \right)}} + \frac{{{\varkappa ^2}\pi _\xi ^2{r^2}\omega _1^2{{\sin }^2}\xi }}{{\left( {\omega _1^2 - \omega _2^2} \right)\left( {\omega _1^2 - \omega _3^2} \right)}}\right.\\
&\left. + \frac{{\varkappa r{\pi _r}\left( {\varkappa {\pi _r}r + 2{\omega _3}} \right)\left( {1 - {r^2}} \right){{\cos }^2}\xi }}{{\omega _1^2 - \omega _3^2}} - \frac{{\varkappa {\pi _\xi }{\omega _1}{r^2}\sin 2\xi }}{{\omega _1^2 - \omega _2^2}} \right]\nonumber\\
 &+ \left[ \frac{{\varkappa \left( {{t_1}\pi _{{\alpha _1}}^4 + {t_2}\pi _{{\alpha _1}}^2\pi _{{\alpha _2}}^2 + {t_3}\pi _{{\alpha _1}}^2\pi _{{\alpha _3}}^2 + {t_4}\pi _{{\alpha _2}}^2\pi _{{\alpha _3}}^2 + {t_5}\pi _{{\alpha _1}}^2 + {t_6}\pi _{{\alpha _2}}^2 + {t_7}\pi _{{\alpha _3}}^2} \right)}}{{\left( {\omega _1^2 - \omega _2^2} \right)\left( {\omega _1^2 - \omega _3^2} \right)}}\right.\nonumber\\
&\left. + \frac{{{\varkappa ^2}\pi _{{\alpha _1}}^2{r^2}\omega _1^2{{\tan }^2}\xi }}{{\left( {\omega _1^2 - \omega _2^2} \right)\left( {\omega _1^2 - \omega _3^2} \right)}} + \frac{{{\varkappa ^2}\pi _{{\alpha _3}}^2{r^2}{{\cos }^2}\xi }}{{\left( {1 - {r^2}} \right)\left( {\omega _1^2 - \omega _3^2} \right)}} \right]\nonumber\ ,\nonumber
\end{align}
\begin{align}\label{IntDefNR2}
{\widetilde I_2} =& {I_2} + \left[ \frac{{\sum\limits_{i = 1}^4 {{s_i}{\varkappa ^i}} }}{{\left( {\omega _2^2 - \omega _1^2} \right)\left( {\omega _2^2 - \omega _3^2} \right)}} + \frac{{{\varkappa ^2}\pi _\xi ^2{r^2}\omega _2^2{{\sin }^2}\xi }}{{\left( {\omega _2^2 - \omega _1^2} \right)\left( {\omega _2^2 - \omega _3^2} \right)}}\right.\\
&\left. + \frac{{\varkappa r{\pi _r}\left( {\varkappa {\pi _r}r + 2{\omega _3}} \right)\left( {1 - {r^2}} \right){{\sin }^2}\xi }}{{\omega _2^2 - \omega _3^2}} - \frac{{\varkappa {\pi _\xi }{\omega _1}{r^2}\sin 2\xi }}{{\omega _2^2 - \omega _1^2}} \right]\nonumber\\
 &+ \left[ \frac{{\varkappa \left( {{t_1}\pi _{{\alpha _1}}^4 + {t_2}\pi _{{\alpha _1}}^2\pi _{{\alpha _2}}^2 + {t_3}\pi _{{\alpha _1}}^2\pi _{{\alpha _3}}^2 + {t_4}\pi _{{\alpha _2}}^2\pi _{{\alpha _3}}^2 + {t_5}\pi _{{\alpha _1}}^2 + {t_6}\pi _{{\alpha _2}}^2 + {t_7}\pi _{{\alpha _3}}^2} \right)}}{{\left( {\omega _2^2 - \omega _1^2} \right)\left( {\omega _2^2 - \omega _3^2} \right)}}\right.\nonumber\\
&\left. + \frac{{{\varkappa ^2}\pi _{{\alpha _1}}^2{r^2}\omega _2^2{{\tan }^2}\xi }}{{\left( {\omega _2^2 - \omega _1^2} \right)\left( {\omega _2^2 - \omega _3^2} \right)}} + \frac{{{\varkappa ^2}\pi _{{\alpha _3}}^2{r^2}{{\sin }^2}\xi }}{{\left( {1 - {r^2}} \right)\left( {\omega _2^2 - \omega _3^2} \right)}} \right]\ ,\nonumber
\end{align}
\begin{align}\label{IntDefNR3}
{\widetilde I_3} =& {I_3} + \left[ \frac{{\sum\limits_{i = 1}^4 {{s_i}{\varkappa ^i}} }}{{\left( {\omega _3^2 - \omega _1^2} \right)\left( {\omega _3^2 - \omega _2^2} \right)}} + \frac{{{\varkappa ^2}\pi _\xi ^2{r^2}\omega _3^2{{\sin }^2}\xi }}{{\left( {\omega _3^2 - \omega _1^2} \right)\left( {\omega _3^2 - \omega _2^2} \right)}}\right.\\
&\left.+ \varkappa r{\pi _r}\left( {\varkappa {\pi _r}r + 2{\omega _3}} \right)\left( {1 - {r^2}} \right)\left( {\frac{{{{\cos }^2}\xi }}{{\omega _3^2 - \omega _1^2}} + \frac{{{{\sin }^2}\xi }}{{\omega _3^2 - \omega _2^2}}} \right) \right]\nonumber\\
&+ \left[ \frac{{\varkappa \left( {{t_1}\pi _{{\alpha _1}}^4 + {t_2}\pi _{{\alpha _1}}^2\pi _{{\alpha _2}}^2 + {t_3}\pi _{{\alpha _1}}^2\pi _{{\alpha _3}}^2 + {t_4}\pi _{{\alpha _2}}^2\pi _{{\alpha _3}}^2 + {t_5}\pi _{{\alpha _1}}^2 + {t_6}\pi _{{\alpha _2}}^2 + {t_7}\pi _{{\alpha _3}}^2} \right)}}{{\left( {\omega _3^2 - \omega _1^2} \right)\left( {\omega _3^2 - \omega _2^2} \right)}}\right.\nonumber\\
&\left. + \frac{{{\varkappa ^2}\pi _{{\alpha _1}}^2{r^2}\omega _3^2{{\tan }^2}\xi }}{{\left( {\omega _3^2 - \omega _1^2} \right)\left( {\omega _3^2 - \omega _2^2} \right)}} + \frac{{{\varkappa ^2}\pi _{{\alpha _3}}^2{r^2}}}{{1 - {r^2}}}\left( {\frac{{{{\cos }^2}\xi }}{{\omega _3^2 - \omega _1^2}} + \frac{{{{\sin }^2}\xi }}{{\omega _3^2 - \omega _2^2}}} \right) \right]\ ,\nonumber
\end{align}
where the undeformed integrals $I_{i}$ are given in equations \eqref{IntUndefNR1}, \eqref{IntUndefNR2} and \eqref{IntUndefNR3}, while the functions $s_{i}$ and $t_{i}$ are defined as
\begin{align}
{s_1} =& \pi _\xi ^2\left( {1 - \frac{1}{{{r^2}}}} \right){\omega _1}\left( {2r{\pi _r}\cos 2\xi  - {\pi _\xi }\sin 2\xi } \right) - {\pi _\xi }\pi _r^2\left( {1 - {r^2}} \right){\omega _1}\sin 2\xi \nonumber\\
  &- {\pi _\xi }{\omega _3}\left( {\omega _1^2 - \omega _2^2} \right)\left( {1 - {r^2}} \right)\sin 2\xi \ ,\nonumber\\
{s_2} =&  - {\pi _\xi }{\pi _r}r\left( {1 - {r^2}} \right)\left( {\omega _1^2 + 2{\omega _3}{\omega _1} - \omega _2^2} \right)\sin 2\xi  - \pi _\xi ^2\pi _r^2\left( {1 - {r^2}} \right){\cos ^2}\xi\nonumber\\ 
&- \pi _\xi ^3\left( {1 - \frac{1}{{{r^2}}}} \right)\left( {2r{\pi _r}\cos \xi  - {\pi _\xi }\sin \xi } \right)\sin \xi  - \pi _\xi ^2\left[ \omega _3^2\left( {1 - {r^2}{{\cos }^2}\xi } \right)\right.\nonumber\\
&\left. + \left( {1 - {r^2}} \right)\left( {\omega _1^2{{\cos }^2}\xi  + \omega _2^2{{\sin }^2}\xi  + 2{\omega _1}{\omega _3}\cos 2\xi } \right) \right]\ ,\nonumber\\
{s_3} =&  - 2{\pi _\xi }\left( {1 - {r^2}} \right)\left( {{\pi _r}r\cos \xi  - {\pi _\xi }\sin \xi } \right)\left[ {\left( {{\omega _1} + {\omega _3}} \right){\pi _\xi }\cos \xi  + {\omega _1}{\pi _r}r\sin \xi } \right]\ ,\nonumber\\
{s_4} =&  - \pi _\xi ^2\left( {1 - {r^2}} \right){\left( {r{\pi _r}\cos \xi  - {\pi _\xi }\sin \xi } \right)^2}\ , \nonumber
\end{align}
\begin{align}
{t_1} =& - \frac{{\varkappa \left( {1 - {r^2}} \right)\left( {1 + {\varkappa ^2}{r^2}} \right){{\sec }^2}\xi {\kern 1pt} \;{{\tan }^2}\xi }}{{{r^2}}}\ ,\nonumber\\
{t_2} =& - \varkappa \left( {1 - {r^2}} \right)\left( {\frac{{{{\sec }^2}\xi }}{{{r^2}}} + {\varkappa ^2}{{\tan }^2}\xi } \right)\ ,\nonumber\\
{t_3} =& - \varkappa \left( {1 + {\varkappa ^2}} \right){\tan ^2}\xi\ ,\nonumber\\
{t_4} =& - \frac{{\varkappa \left( {1 + {\varkappa ^2}} \right){{\cot }^2}\xi }}{{1 - {r^2}}}\ ,\nonumber\\
{t_5} =& \frac{{2\left( {1 - {r^2}} \right)\tan \xi }}{r}\left( \frac{{\varkappa \left( {1 + {\varkappa ^2}{r^2}} \right){\pi _\xi }(r{\pi _r} - {\pi _\xi }\tan \xi )}}{r} + {\omega _1}{\pi _r}\left( {1 + {\varkappa ^{\rm{2}}}{{r}^{\rm{2}}}} \right)\tan \xi \right.\nonumber\\
&\left. + \frac{{{\pi _\xi }\left( {\left( {1 + {\varkappa ^{\rm{2}}}{{r}^{\rm{2}}}} \right){\omega _1}{ + }{\varkappa ^{\rm{2}}}{\omega _3}{{r}^{\rm{2}}}} \right)}}{{r}} \right) - \varkappa \left( {\left( {\omega _2^2 - 2{\omega _1}{\omega _3}} \right)\left( {1 - {r^2}} \right) + \omega _3^2} \right){\tan ^2}\xi\ ,\nonumber\\
{t_6} =& \frac{{\left( {1 - {r^2}} \right)}}{{{r^2}}}\left[  - \varkappa \left( {1 + {\varkappa ^2}{r^2}} \right){{(r{\pi _r}\cot \xi  - {\pi _\xi })}^2} - 2\left( {{\omega _1} + ({\omega _1} + {\omega _3}){\varkappa ^2}{r^2}} \right) (r{\pi _r}\cot \xi  - {\pi _\xi })\cot \xi\right.\nonumber\\
&\left. - \varkappa {{({\omega _1} + {\omega _3})}^2}{r^2}{{\cot }^2}\xi \right] ,\nonumber\\
{t_7} =& - \frac{{{\pi _\xi }\left( {1 + {\varkappa ^2}} \right)}}{{1 - {r^2}}}\left( {{\omega _1}{r^2}\sin 2\xi  + \varkappa {\kern 1pt} {\pi _\xi }\left( {1 - {r^2}{{\sin }^2}\xi } \right)} \right)\nonumber\ .
\end{align}
Just as in the undeformed case, these expressions satisfy all the properties explained in Section \ref{Defsystems}, but this time with the canonical Poisson brackets in the $(r,\xi)$ coordinates, instead of the Dirac brackets used in the $x_i$ coordinates.

\end{document}